\documentclass{article}
\usepackage[english]{babel}
\usepackage[utf8x]{inputenc}
\usepackage{geometry}
\usepackage{amsmath}
\usepackage{subfig}
\usepackage{amsfonts}
\usepackage{amssymb}
\usepackage{graphicx}
\usepackage{siunitx}
\usepackage{chngpage}
\usepackage[left]{lineno}%
\providecommand{\U}[1]{\protect\rule{.1in}{.1in}}
\title{Measurement of Reconstructed Charged Particle Multiplicities of Neutrino Interactions in MicroBooNE\\{\normalsize {Talk presented at the APS Division of Particles and Fields Meeting (DPF 2017), July 31-August 4, 2017, Fermilab. C170731}}}
\author{Aleena Rafique, for the MicroBooNE Collaboration\\ {\normalsize {Kansas State University}}}
\date{\Large{\today}}
\begin{document}
\maketitle
\begin{abstract}
We compare the observed charged particle multiplicity distributions
in the MicroBooNE liquid argon time projection chamber from neutrino
interactions in a restricted final state phase space to predictions of this
distribution from several GENIE models. The measurement uses a data sample
consisting of neutrino interactions with a final state muon candidate fully
contained within the MicroBooNE detector. \ These data were collected in
2015-2016 with the Fermilab Booster Neutrino Beam (BNB), which has an average
neutrino energy of $800$ MeV, using an exposure corresponding to $%
5.0\times10^{19}$ protons-on-target. \ The analysis employs fully automatic
event selection and charged particle track reconstruction and uses a
data-driven technique to determine the contribution to each multiplicity bin from neutrino interactions and cosmic-induced backgrounds. The restricted phase space employed makes the measurement most
sensitive to the higher-energy charged particles expected from primary
neutrino-argon collisions and less sensitive to lower energy protons
expected to be produced in final state interactions of collision products
with the target argon nucleus.
\end{abstract}

\tableofcontents

\newpage\newpage

\section{Introduction}

\subsection{Observed Charged Particle Multiplicity Distribution}

Neutrino interactions in the MicroBooNE detector~\cite{MicroBooNE detector}
produce charged particles that can be reconstructed as tracks in the liquid
argon medium of the active\ time projection chamber (TPC) volume of the detector. The charged
particle multiplicity, or number of primary charged stable particles $n$, is
a simple observable characterizing final states in high-energy-collision
processes, including neutrino interactions. We note that in MicroBooNE the observable charged particle multiplicity corresponds to that of charged particles exiting the target nucleus participating in the neutrino interaction. The charged particle
multiplicity distribution (CPMD) comprise the set of probabilities, $P_{n}$%
, associated with producing $n$ charged particles in a neutrino event, either in
full phase space or in restricted phase space domains.

Determination of the CPMD at Fermilab Booster Neutrino Beam (BNB)~\cite{BNB
reference} neutrino energies naturally fits into the modern strategy~\cite%
{Sam XC Review} of presenting neutrino interaction measurements in the form
of directly observable quantities. \ The CPMD measurements expand the
empirical knowledge of neutrino-argon scattering that will be required by
the DUNE experiment~\cite{DUNE reference}. As
physical observable, the CPMD can also be used to test neutrino cross-section models, or
particular tunes of generators such as GENIE~\cite{GENIE reference}. \ These
models are typically constructed from a set of exclusive cross section
channels, and the inclusive character of CPMD thus provides independent
checks.

This note describes a preliminary test of several variants of GENIE against
the \emph{observed} charged particle multiplicity distribution (observed
CPMD) in MicroBooNE data collected during the 2015-2016 time period at the
Fermilab BNB. \ By \textquotedblleft observed\textquotedblright , we mean
the probability, after application of certain detector selection
requirements, of observing a neutrino interaction with $n$ charged tracks. We present this relative to the probability of observing a neutrino interaction with at
least one charged track:%
\begin{equation}
O_{n}=\frac{N_{\text{obs,}n}}{\sum\limits_{m=1}^{M_{\text{obs},\max }}N_{%
\text{obs,}m}},
\end{equation}%
where $N_{\text{obs,}n}$ is the number of neutrino interaction events with $n
$ observed tracks, and $M_{\text{obs},\max }$ is the maximum observed value
for $N_{\text{obs,}n}$. \ At BNB energies, $M_{\text{obs},\max }\approx 5$.
Our analysis requires at least one of the charged tracks to be consistent
with a muon; hence the $O_{n}$ are effectively observed CPMD for $\nu _{\mu }
$ charged current ($\nu _{\mu }$ CC) interactions. \ The muon candidate is
included in the charged particle multiplicity, and all events thus have $%
n\geq 1$. \ 

The values for $O_{n}$ depend on cross sections for producing a multiplicity 
$n$, $\sigma _{CC,n}$ as well as the BNB neutrino flux and detector
acceptance and efficiency:%
\begin{equation}
N_{\text{obs,}n}=\sum_{n^{\prime
}}\int dE_{\nu }\Phi _{\nu }\left( E_{\nu }\right) \int d\Pi _{n^{\prime }}%
\frac{d\sigma _{CC,n^{\prime }}\left( E_{\nu },\Omega \right) }{d\Pi
_{n^{\prime }}}\epsilon _{n,n^{\prime }}\left( E_{\nu },\Pi _{n^{\prime
}}\right) ,  \label{N-observed}
\end{equation}%
where $E_{\nu }$ is the neutrino energy, $\Phi _{\nu }\left( E_{\nu }\right) 
$ is the neutrino flux, $d\Pi _{n}$ represents the $n$-particle final state phase space, $%
\epsilon _{n,n^{\prime }}\left( E_{\nu },\Omega \right) $ is an acceptance
and efficiency matrix that gives the probability that an $n^{\prime }$ charged
particle final state produced in phase space element $d\Pi _{n^{\prime }}$
is observed as an $n$-particle final state in the detector, and $d\sigma
_{CC,n^{\prime }}\left( E_{\nu },\Omega \right) /d\Pi _{n}$ are the
differential cross sections for producing a multiplicity $n^{\prime }$. The data are not corrected for $\nu_\mu$ NC, $\nu_{e}$, $\bar\nu_e$, or $\bar{\nu}_\mu$ backgrounds. These backgrounds, in total, are expected to be less than 15\% of the final sample. An assumption is made that the Monte Carlo simulation adequately describes these non $\nu_\mu$ CC backgrounds. \ In
practice we obtain the $O_{n}$ directly from data and compare these to
values derived from evaluating Eq. \ref{N-observed} using a Monte Carlo
simulation that includes GENIE\ neutrino interactions event generators
coupled to detailed GEANT-based~\cite{Geant4 reference} models of the
Fermilab BNB and the MicroBooNE detector.

The observed CPMD has several desirable attractive attributes. The $\sigma
_{CC,n}$ are all relatively large up to $n\lesssim 4$ so only modest
statistics are required. Approximately 1100 reconstructed  neutrino events contribute to this analysis. \ Only minimal kinematic properties of the final
state are imposed (track definition implies an effective minimum kinetic
energy), and complexities associated with particle identification are
avoided. \ Similarly, restricting the analysis to tracks avoids
issues associated with electromagnetic shower reconstruction. At the same time, the observed
CPMD exploits much of the power of the liquid argon TPC in identifying and
characterizing complex neutrino interactions. Observed charged particles multiplicities obtained from the data can help provide a more stringent test of event generators and can be used to constrain them. \ Finally, as the observed
CPMD are ratios, they would be expected to have reduced sensitivity to
systematic uncertainties associated with flux and detector systematics compared to
absolute cross section measurements.

A disadvantage of observed CPMD is its lack of portability. One must have
access to the full MicroBooNE\ simulation suite to use the $O_{n}$ to test
models. In a future publication we will \textquotedblleft
unfold\textquotedblright\ the observed CPMD, correcting for efficiency and
flux effects, and provide values of $P_{n}$ corresponding to well defined
kinetic energy thresholds.

\subsection{MicroBooNE\ Detector and the Booster Neutrino Beam}

\begin{figure}[ptb]
\centering
\includegraphics[width=0.7\textwidth]{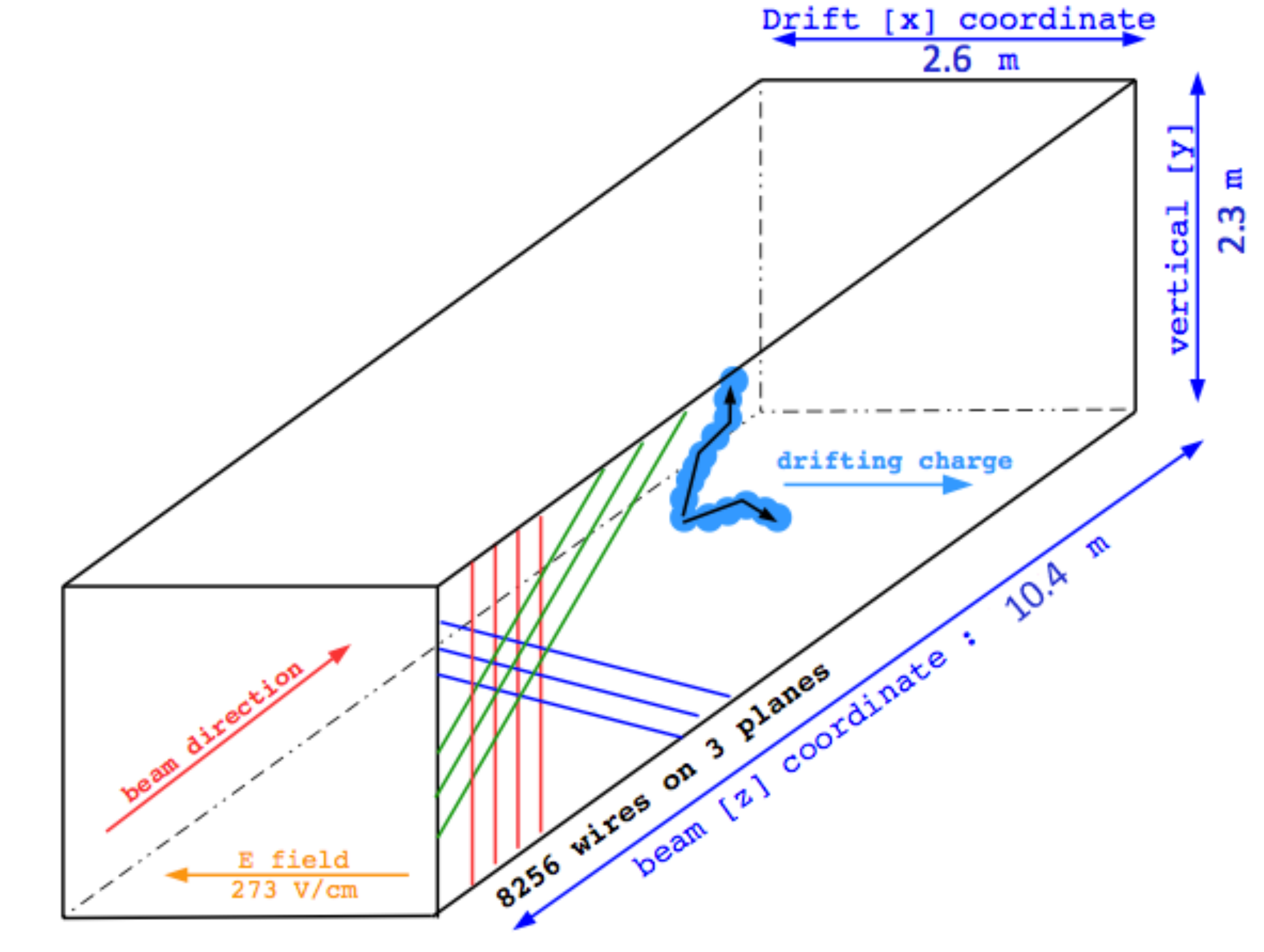} 
\caption{A schematic of the MicroBooNE TPC showing the coordinate
system and wire planes.}
\label{img:MicroBooNE picture}
\end{figure}

\begin{figure}[tbp]
\centering
\includegraphics[width=0.7\textwidth]{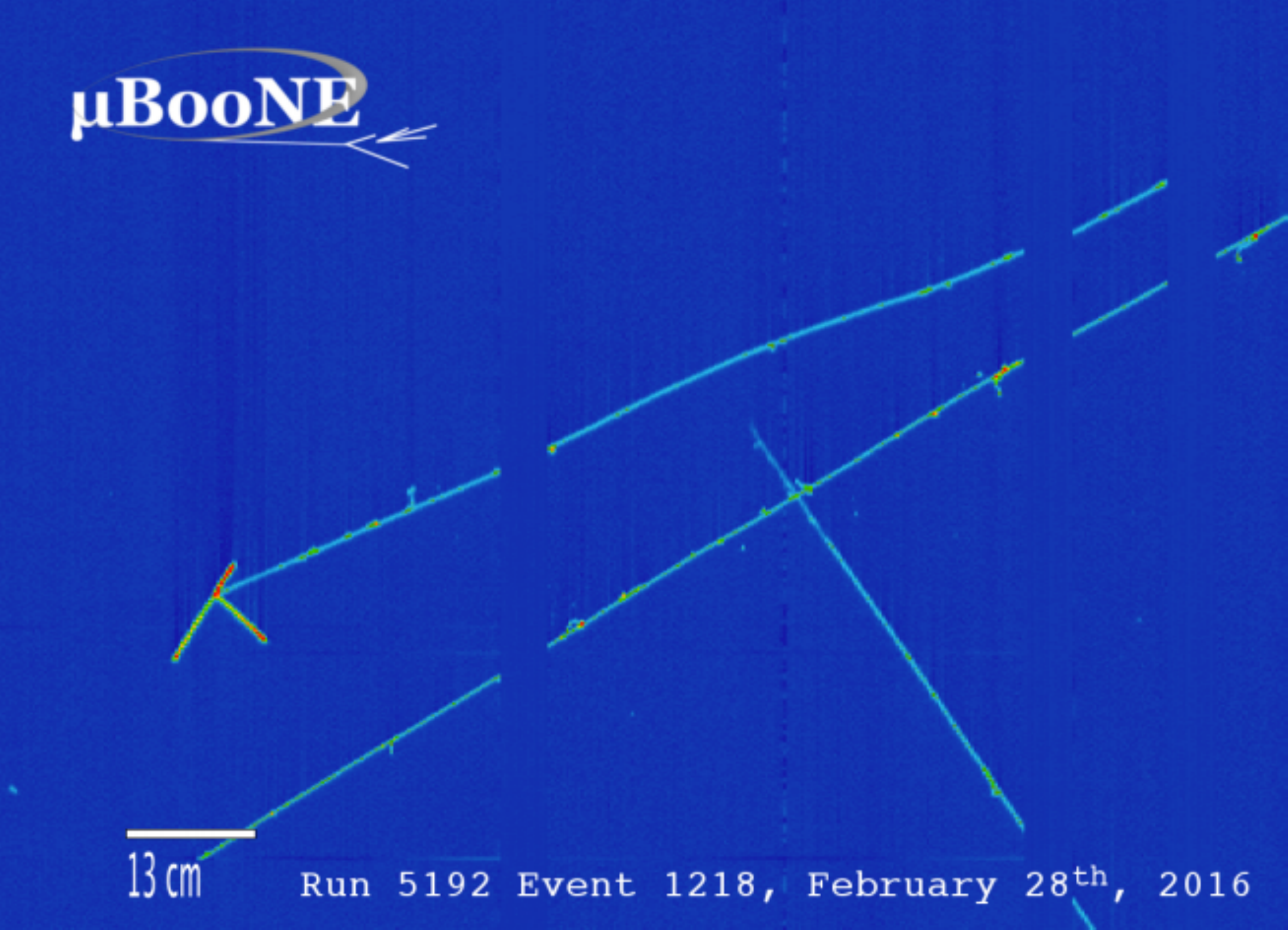} 
\caption{Event display showing raw data for a region of the collection plane
associated with a candidate high-multiplicity neutrino event. Wire-number is
represented on the horizontal axis, and time on the vertical. Color is
associated with the charge deposition on each wire.}
\label{img:Event display}
\end{figure}

The MicroBooNE detector (Figure~\ref{img:MicroBooNE picture}) is a
liquid-argon time projection chamber TPC installed on the Fermilab BNB.
MicroBooNE is a high-resolution detector designed to be able to accurately
identify low energy neutrino interactions. It began collecting neutrino beam
data in October of 2015. \ Figure~\ref{img:Event display} shows an image of a high multiplicity event from MicroBooNE\ data.

The MicroBooNE TPC has an active mass of about 90 tons of liquid argon. It is 10.4 meters long in the beam direction, 2.3 meters tall, and 2.6 meters in the
electron drift direction. Electrons require  2.3 ms  to drift across the
full width of the TPC at the -70 kV operating voltage. Events are read out
on three anode wire planes with 3 mm spacing between wires. Drifting
electrons pass through the first two wire planes, which are oriented at $\pm
60$ degrees relative to vertical, producing bipolar induction signals. \ The
third \textquotedblleft collection plane\textquotedblright\ (CP) has its
wires oriented vertically and collects the charge of the drifting electrons
in the form of a unipolar signal. \ The MicroBooNE readout electronics
allow for measurement of both the time and charge created by drifting
electrons on each wire.

While all three anode planes are used for track reconstruction, the CP
provides the best signal-to-noise performance and charge resolution. \ The
analysis presented here excludes regions of the detector that have
non-functional CP channels. It also imposes requirements on the minimum number of CP hits$-$electric current pulses processed through noise filtering ~\cite{Noise Filtering}, deconvolution, and calibration operations$-$ associated with the reconstructed tracks.  All charged particle track candidates are required to have at least 20 CP hits, and the longest muon track candidate is required to have at least 80 CP hits. \ Furthermore, as described in Sec. %
\ref{sig_extraction}, we use two discriminants to extract the neutrino
interaction and cosmic ray background contributions to our data sample that
are based on CP hits.

A light collection system consisting of 32 8-inch photomultiplier tubes with
nanosecond timing resolution enables precise determination of the initial
time of the neutrino interaction, which crucially aids in the reduction of
cosmic ray backgrounds.

The BNB employs protons from the Fermilab Booster synchrotron impinging on a
beryllium target. The proton beam has a kinetic energy of 8 GeV, a
repetition rate of up to 5 Hz, and is capable of producing $5\times10^{12}$
protons-per-spill. Secondary pions and kaons decay producing neutrinos with
an average energy of $800$ MeV. The estimated BNB flux is shown in Fig. \ref%
{img:BNB flux figure}. MicroBooNE received $3.6\times10^{20}$
protons-on-target in its first year of running from fall 2015 through summer
2016. \ This analysis uses a fraction of that data corresponding to $%
5.0\times10^{19}$ protons-on-target.

\begin{figure}[ptb]
\centering
\includegraphics[width=0.6\textwidth]{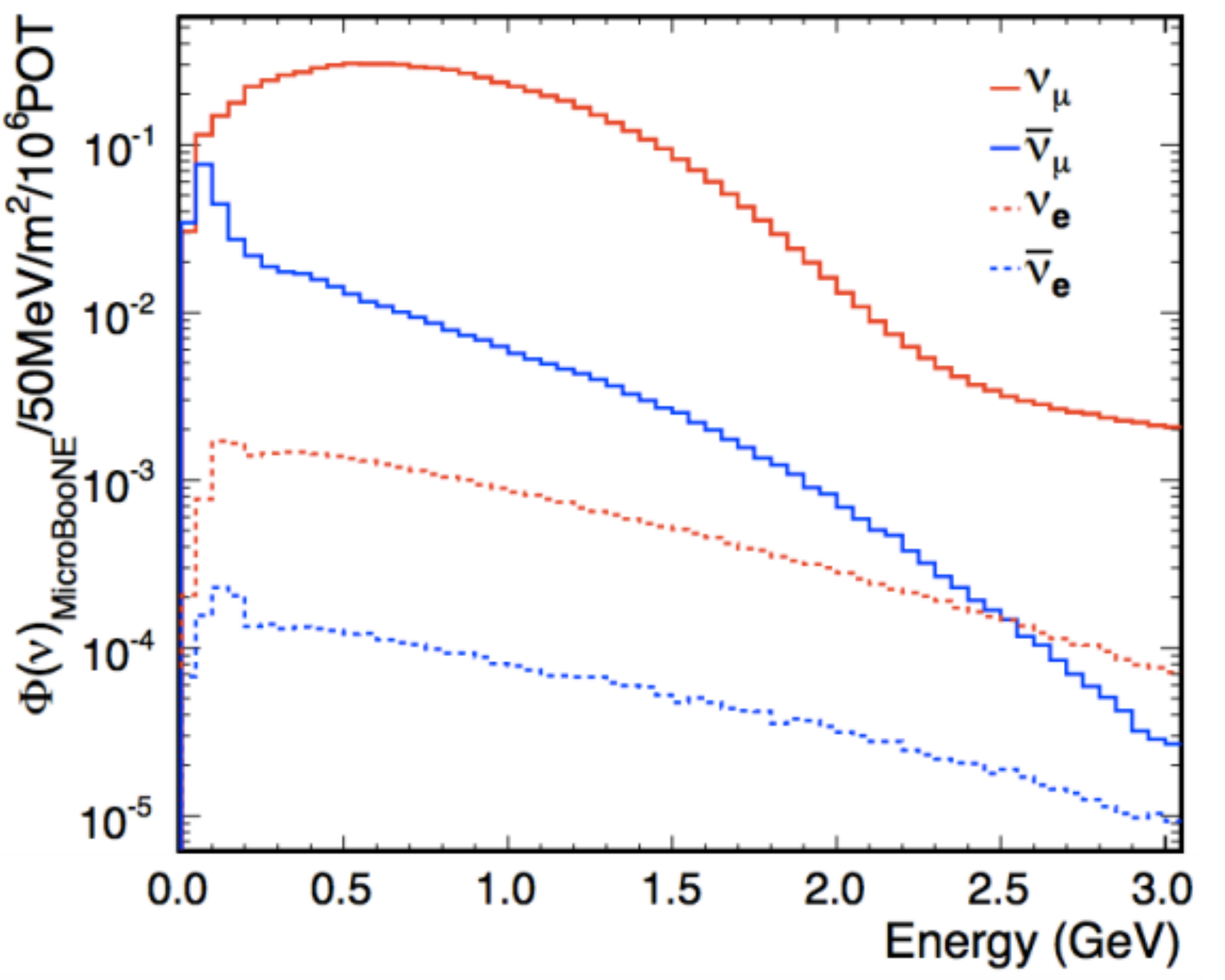} 
\caption{Booster Neutrino Beam flux at the MicroBooNE location for each neutrino
flavor.}
\label{img:BNB flux figure}
\end{figure}

\section{Data and Simulation Samples}

\subsection{Data}
This analysis uses two data samples:
\begin{itemize}
\item The  \textquotedblleft on-beam data\textquotedblright, which
is taken only during periods when a beam spill from the BNB is actually
sent. The on-beam data used were recorded from February
to April  2016 using data taken in runs in which the BNB and all detector systems
functioned well~\cite{DocDB5986-Run1DQ-PUB}. \ This sample comprises about $%
15\%$ of the total neutrino data collected by MicroBooNE\ in its first
running period. \ The remaining data will be added in the future.
\item The \textquotedblleft off-beam data\textquotedblright, which is taken
with the exact same trigger settings as the on-beam data, but during periods
when no beam was received. The off-beam data were collected from February to October 2016. 
\end{itemize}

\subsection{Simulation}

The LArSoft software framework~\cite{Larsoft reference} is used for
processing data events and Monte Carlo simulation events in the same way.
Five simulation samples are used in this analysis: 
\begin{itemize}
\item Neutrino interactions generated with default GENIE (\textquotedblleft BNB-only MC\textquotedblright ).
\item CORSIKA~\cite%
{Corsika} cosmic ray (CR) simulation (\textquotedblleft CR-only MC\textquotedblright ).
\item Neutrino interactions simulated with a default GENIE model overlaid with CORSIKA CR events (\textquotedblleft BNB+Cosmic default MC\textquotedblright ). 
\item BNB+Cosmic generated with the GENIE\ implementation of the Meson
Exchange Currents model (\textquotedblleft BNB+Cosmic with
MEC\textquotedblright ).
\item BNB+Cosmic with the GENIE\ implementation of
the Transverse Enhancement Model (\textquotedblleft BNB+Cosmic with
TEM\textquotedblright ).
\end{itemize}

The generator stage employs GENIE version 2.8.6~\cite{GENIE reference} with
overlaid simulated CR backgrounds using CORSIKA version v7.4003~\cite{Corsika}%
. Simulated secondary particle propagation utilizes GEANT version v4.9.6.p04d~\cite{Geant4 reference}, and detector response simulation employs LArSoft
version v4.36.00. All GENIE samples were processed with the same GEANT and
LArSoft versions for detector simulation and reconstruction. These samples
thus allow for relative comparison of different GENIE models to the data.

\section{Event Selection and Signal Extraction}

In this analysis, an event selection starts by applying a filter algorithm
(\textquotedblleft $\nu _{\mu }$ CC filter\textquotedblright ) that selects 
\emph{$\nu _{\mu }$ }candidates~\cite{DocDB5982-CC-XC-PUB}. A set of
conditions are then applied to ensure a high quality muon track candidate
(\textquotedblleft good track filter"). Finally, a \textquotedblleft muon
directionality classifier\textquotedblright\ is implemented that categorizes
the final selected events into four sub-samples for the purpose of CR
background estimation.

\subsection{$\protect\nu _{\protect\mu }$ CC Filter}

This filter requires events to contain a track candidate that has a minimum
reconstructed length of 75 cm that is fully contained within the fiducial
volume of the detector. \ The light associated with the track candidate must arrive in time coincidence with the beam spill time window, and the candidate must generate a pattern of PMT hits consistent with those expected for a muon track produced in coincidence
with the beam arrival. Further details of this selection can be found
in Ref.~\cite{DocDB5982-CC-XC-PUB}. Considerable CR backgrounds remain after this first stage filter, with signal/background $\simeq $ 1/1.  Simple CR background subtraction techniques prove to be insufficient. Later sections describe a new method developed for this analysis to extract neutrino interaction contributions to the observed CPMD.

\subsection{Good Track Filter}

Pre-selected events then pass through another filter that imposes further
quality conditions on track candidates. Start and end points of the
candidate muon must lie in detector regions with well-functioning CP wires. Furthermore, the candidate
muon track must have at least 80 hits in the collection wire plane, and it
must not have significant wire gaps in the start and end 20 CP-hit segments
used in the pulse height (PH) test (Sec. \ref{PH}) and the multiple Coulomb
scattering (MCS) test\emph{\ }(Section \ref{MCS}).

Events passing the \emph{$\nu _{\mu }$ }CC filter and the good data filter
comprise the final data sample. Table \ref{tab:passingrates} lists the event
passing rates for the on-beam data, off-beam data, and the BNB+Cosmic MC
samples at different steps of the event selection. 
\begin{table}[tbp]
\caption{Passing rates for event selection criteria applied to on-beam data,
off-beam data, and BNB+Cosmic MC samples. Numbers are absolute event counts.
Quantities in parentheses give the relative passing rate with respect to the
step before (first percentage) and the absolute passing rate with respect to
the starting sample (second percentage).}
\label{tab:passingrates}\centering
\begin{adjustwidth}{-0.8cm}{}
\renewcommand{\tabcolsep}{1pt}
\begin{tabular}
[c]{l|c|c|c|c|c|c}

& \multicolumn{2}{|c}{\textbf{On-beam Data}} & \multicolumn{2}{|c}{\textbf{Off-beam Data}}& \multicolumn{2}{|c}{\textbf{BNB+Cosmic Default MC}}\\\
\textbf{Selection Cuts}& \textbf{events} &\textbf{passing rates} &\textbf{events} & \textbf{passing rates} & \textbf{events} & \textbf{passing rates} \\\hline
Generated events & 547,616 & & 2,954,586 &  & 188,880 & \\
$\nu_{\mu}$ CC filtered events & 4075 & (0.7\%/0.7\%)& 14340 & (0.5\%/0.5\%)& 7106 & (3.8\%/3.8\%)\\
Events passing dead region cut & 2577 & (63.2\%/0.5\%)& 8206 & (57.2\%/0.3\%)& 4993 & (70.3\%/2.6\%) \\
Events with $\geq 80$ CP hits & 2059 & (80.0\%/0.4\%)& 5608 & (68.3\%/0.2\%)& 4591 & (91.9\%/2.4\%) \\ 
Events passing segments gap cuts & 1921 & (93.3\%/0.4\%)& 5267 & (93.9\%/0.2\%)& 4209 & (91.7\%/2.2\%)\\
\end{tabular}
 \end{adjustwidth}
\end{table}

\subsection{Muon Directionality Classifier}

Events satisfying all of the selection criteria are further categorized into
the four sub samples based on whether they pass or fail the PH test and the
MCS tests described below. These are the tests of the direction of the candidate
muon that can be used to separate neutrino signal and CR background contributions
in the sample. Table \ref{tab:acceptance_rates} lists the event selection
rates for the on-beam data, off-beam data, and the BNB+Cosmic default MC samples in
each sub sample. 
\begin{table}[tbp]
\caption{Final event samples from the on-beam data, off-beam data, and
BNB+Cosmic MC samples. Numbers are absolute event counts. The percentages
correspond to the fraction of events in each test category.}
\label{tab:acceptance_rates}\centering
\renewcommand{\tabcolsep}{1pt} 
\begin{tabular}{c|c|c|c|c|c|c}
\textbf{Sub samples} & \multicolumn{2}{|c}{\textbf{On-beam Data}} & \multicolumn{2}{|c}{
\textbf{Off-beam Data}} & \multicolumn{2}{|c}{\textbf{BNB+Cosmic Default MC}} \\ 
\textbf{PH, MCS} & \textbf{events} & \textbf{acceptance rates} & \textbf{events} & \textbf{acceptance rates} & \textbf{events} & \textbf{acceptance rates} \\ \hline
pass, pass & 847 & (44\%) & 1263 & (24\%) & 2629 & (62\%) \\ 
pass, fail & 367 & (19\%) & 1087 & (21\%) & 737 & (18\%) \\ 
fail, pass & 321 & (17\%) & 1141 & (22\%) & 440 & (10\%) \\ 
fail, fail & 386 & (20\%) & 1776 & (34\%) & 403 & (10\%) \\ 
\end{tabular}%
\end{table}

\subsubsection{Pulse Height Test}

\label{PH} A neutrino-induced muon from a CC event will create a track that
usually travels in the beam direction (``upstream" to ``downstream") and has an increasing rate of energy loss as one moves downstream along the track. \ Selection criteria can be applied to
pick tracks that satisfy this expectation in order to suppress cosmic
backgrounds.

We take into account the expected behavior of the rate of restricted energy loss~\cite{Particle Data Group}, $dE/dx_{R}$%
, with the following procedure:

\begin{itemize}
\item Compute the truncated mean of the charge deposited in $20$ consecutive CP hits, $\left\langle PH\right\rangle _{U}$, starting $10$ hits
away from the upstream end of the muon track that is taken as a proxy for the
upstream restricted energy loss. \ The truncated mean is formed by
taking the average of the $20$ PH after removing individual PH that do not
lie within the range of $20\%-200\%$ of the average~\cite{CCFR 20-200}:%
\begin{equation}
\left\langle PH\right\rangle _{U}=\frac{\sum\limits_{n=11}^{n=30}PH_{n}%
\left( 0.2\left\langle PH\right\rangle <PH_{n}<2.0\left\langle
PH\right\rangle \right) }{\sum\limits_{n=11}^{n=30}\left( 0.2\left\langle
PH\right\rangle <PH_{n}<2.0\left\langle PH\right\rangle \right) },
\end{equation}%
which can be determined iteratively with an initial approximation that $%
\left\langle PH\right\rangle $ is the arithmetic average. \ Use of the
truncated mean PH rather than the average PH minimizes effects of large
energy loss fluctuations. \ 

\item Form a similar quantity from $20$ consecutive CP hits that end 10 CP
hits away from the downstream end of the track, $\left\langle PH\right\rangle _{D}$.

\item Form the test $p=\left\langle PH\right\rangle _{U}<\left\langle
PH\right\rangle _{D}$. \ Muons from $\nu _{\mu }$ CC interactions will pass
this test with a probability $P\left( PH\right) $. \ Muons from CR
background can be characterized by the probability that they fail the
negation of the test $\bar{p}=\left\langle PH\right\rangle _{U}>\left\langle
PH\right\rangle _{D}$, denoted as $Q\left( PH\right) .$
\end{itemize}

\begin{figure}[tbp]
\centering
\includegraphics[width=0.7\textwidth]{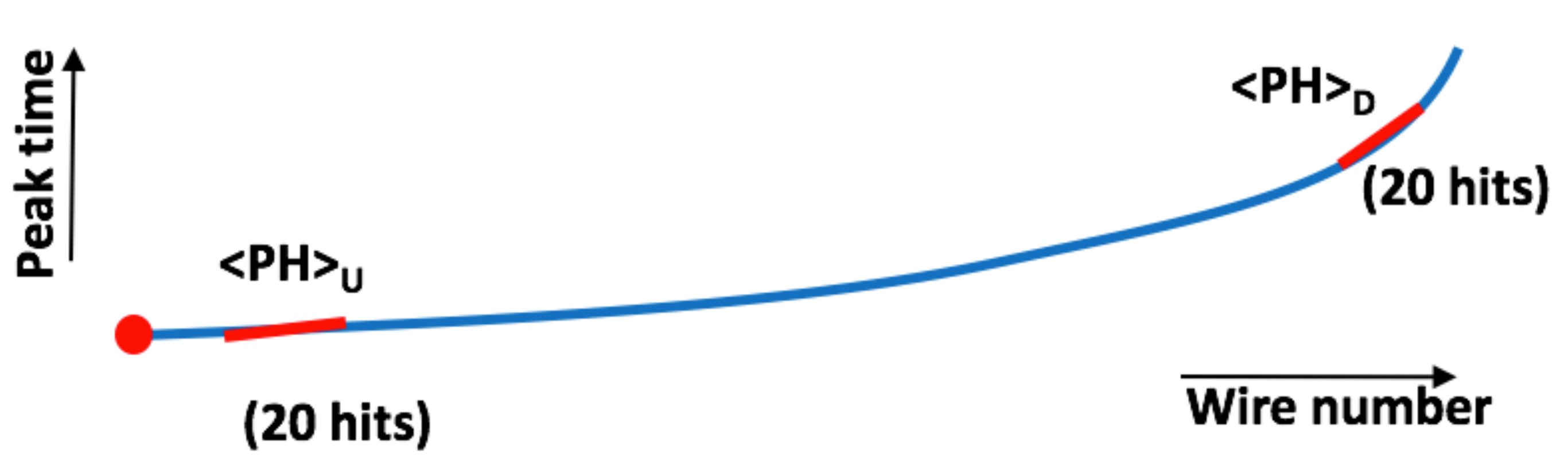} 
\caption{Diagram showing PH test for a candidate muon track.}
\label{img:PH test}
\end{figure}

\begin{figure}[tbp]
\centering
\includegraphics[width=0.7\textwidth]{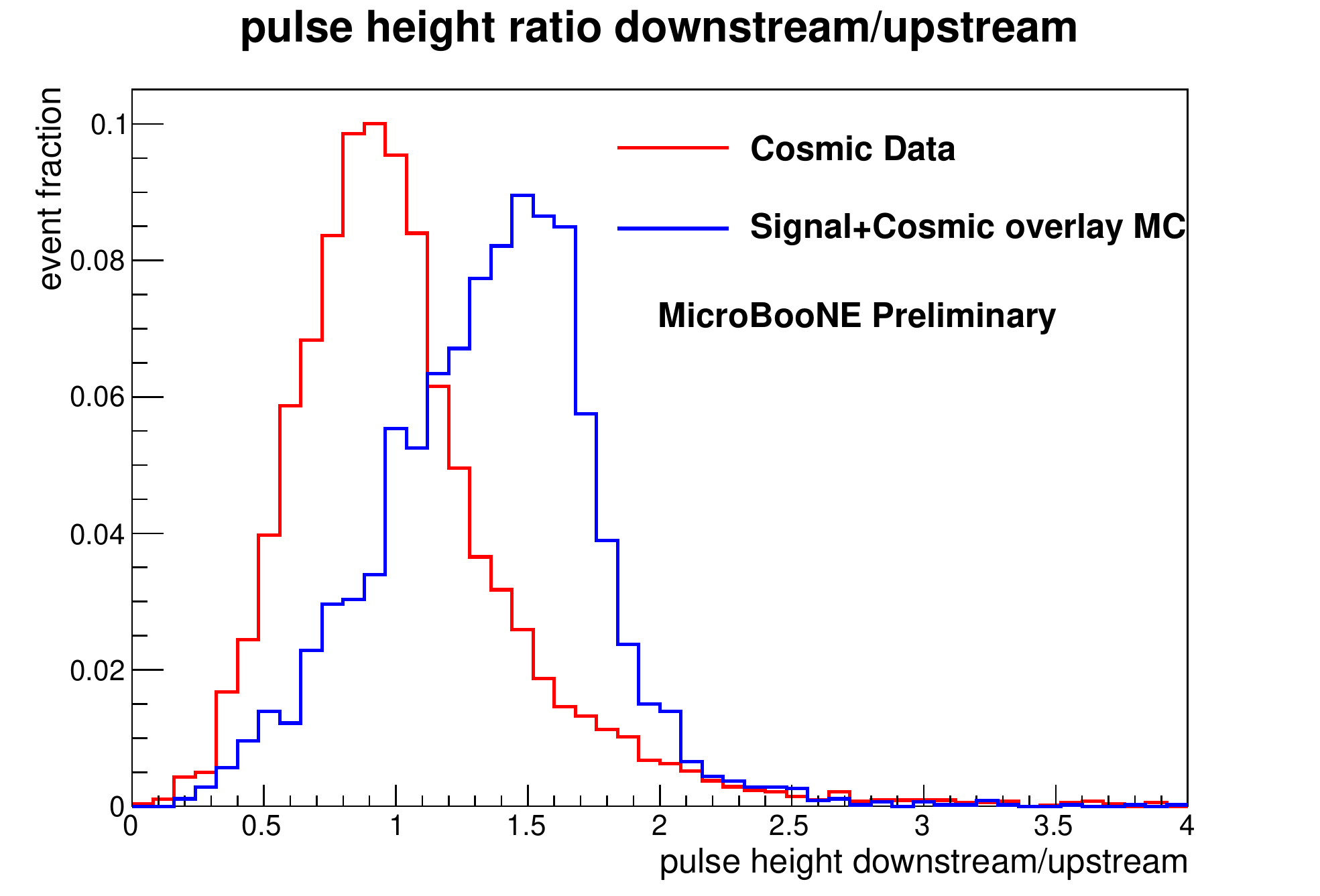} 
\caption{Pulse height (PH) downstream to upstream ratio $\left\langle
PH\right\rangle _{D}/\left\langle PH\right\rangle _{U}$.}
\label{img:PH_ratio}
\end{figure}

A visual diagram for the PH test is shown in Figure \ref{img:PH test}.
Figure \ref{img:PH_ratio} presents the pulse height downstream to upstream
ratio distribution for BNB+Cosmic default MC (signal+cosmic background) and off-beam
data (cosmic background only). We observe that PH ratio for the signal
significantly dominates over the background for values $>$1, which was
the cut value chosen for the signal-enhanced sample used in the analysis.

\subsubsection{MCS Test}

\label{MCS}
A neutrino-induced muon from a CC event will create a track that usually travels in the beam direction (``upstream" to ``downstream") and has an increasing degree of scatter about a nominal straight line trajectory as one moves downstream along the track. \ Selection criteria can be applied to
pick tracks that satisfy this expectation in order to suppress cosmic
backgrounds.

The expected MCS behavior is taken into account by an independent test with the following procedure:

\begin{itemize}
\item Take three 20 CP-hit long track segments at the upstream, downstream,
and geometric center of the track. \ The upstream and downstream segments
are displaced by 10 CP hits from the upstream and downstream ends of the
track, respectively.

\item Perform a simple linear least squares fit of hit time vs. (wire) position
using the $20$ contiguous CP hits at the upstream end of the track. \ Denote
the determined line as $L_{U}$. Perform a similar fit using the $20$ CP hits at the downstream end of the
track. \ Denote the determined line as $L_{D}.$ Finally perform one more similar fit from the $20$ CP hits located about the
geometric center of the track. Denote this line as $L_{M}$.

\item Compare the hit time (drift direction `$x$') predicted at the geometric center of the track, $%
t_{C}$, by $L_{M}$, which uses hits about the geometric center, to the time
predicted at the geometric center of the track by the projection of $L_{U}$
from the beginning of the track: 
\begin{equation}
\Delta t_{UM}=\left\vert t_{C}\left( L_{U}\right) -t_{C}\left( L_{M}\right)
\right\vert .
\end{equation}

\item Repeat the process except compare $t_{C}$ from $L_{M}$ to the time
predicted at the geometric center of the track by the projection of $L_{D}$
from the end of the track:%
\begin{equation}
\Delta t_{DM}=\left\vert t_{C}\left( L_{D}\right) -t_{C}\left( L_{M}\right)
\right\vert .
\end{equation}

\item Form the test $q=\Delta t_{UM}<\Delta t_{DM}$. \ Since MCS should
become, on average, more pronounced along the downstream end of the track as
the momentum decreases, this provides a second directional test on the muon
track candidate. \ Muons from $\nu _{\mu }$ CC interactions will pass this
test with a probability $P\left( MCS\right) $. \ Muons from CR background
tests can be characterized by the probability that they fail the negation of
the test $\bar{q}=\Delta t_{UM}>\Delta t_{DM}$, denoted as $%
Q\left( MCS\right) .$
\end{itemize}

\begin{figure}[tbp]
\centering
\includegraphics[width=0.7\textwidth]{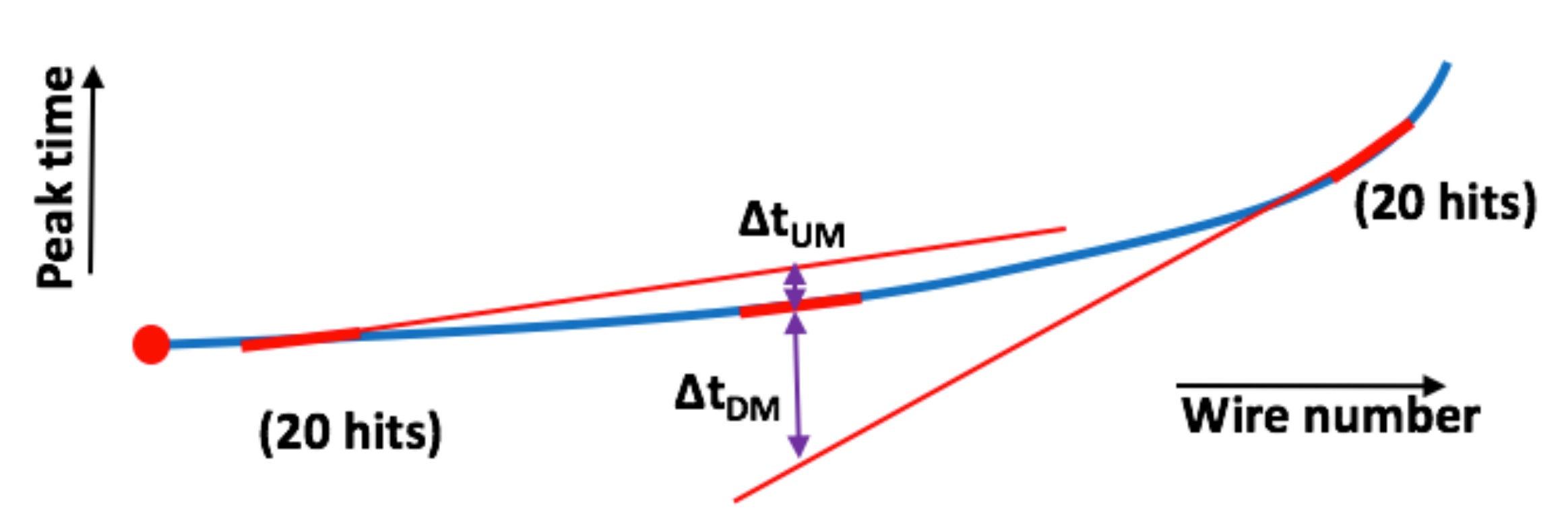} 
\caption{Diagram of MCS directionality test for a candidate muon track.}
\label{img:MCS test}
\end{figure}

\begin{figure}[tbp]
\centering
\includegraphics[width=0.7\textwidth]{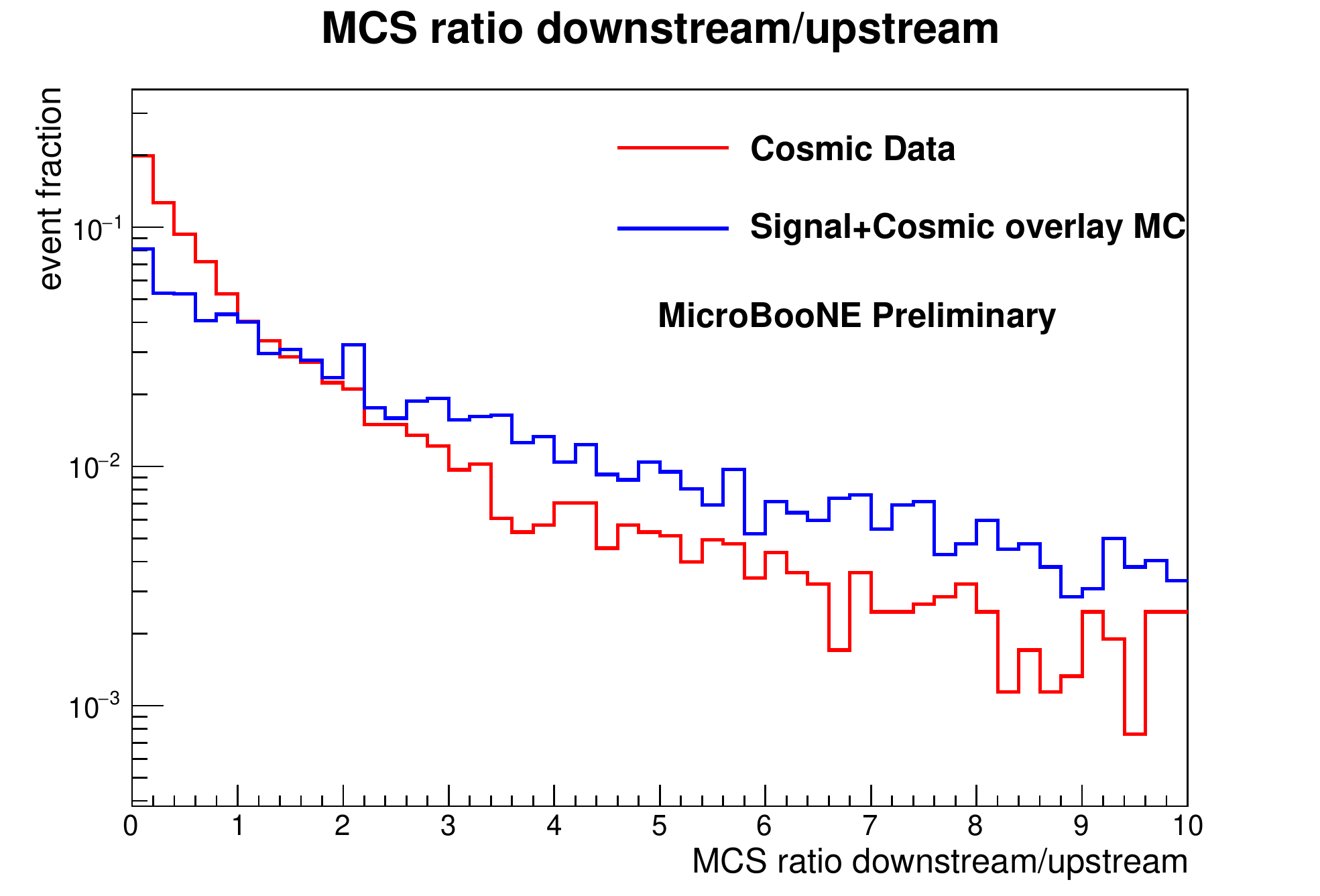} 
\caption{Multiple Coulomb scattering (MCS) downstream to upstream ratio $%
\Delta t_{DM}/\Delta t_{UM}$. }
\label{img:MCS_ratio}
\end{figure}

A visual diagram for the MCS test is shown in Figure \ref{img:MCS test}.
Figure \ref{img:MCS_ratio} presents the MCS downstream to upstream ratio $%
\Delta t_{DM}/\Delta t_{UM}$ distribution for BNB+Cosmic default MC (signal+cosmic
background) and off-beam data (cosmic background only). We observe that MCS
ratio for the signal dominates over the background for values $>$1, which
was the cut value chosen for the signal-enhanced sample used in the analysis.\\

\section{Expectations for Observed Charged Particle Multiplicity Distribution\ }

All $\nu _{\mu }$ CC events inherently have a muon and at least one charged
hadron at the initial neutrino interaction vertex. In order to mitigate
backgrounds from cosmic ray (CR) interactions, which are significant due
MicroBooNE's placement at the earth's surface, muon track candidates must
produce a visible track with a physical length of at least 75 cm in argon
that has all its TPC hits contained within a detector fiducial volume defined to be the liquid argon volume within the TPC that is greater than 5 cm from any TPC boundary. \
The candidate muon containment requirement limits its energy to be $\lesssim 1.2$ GeV depending on the muon scattering angle. This results in a sample biased towards relatively higher
inelasticity $y=\nu /E_{\nu }$, in the rest frame of the hadronic system, with $\nu$ the energy transferred from the
neutrino to the hadronic system in the collision.

At BNB energies, the nominally dominant charged particle multiplicities at
the neutrino interaction point are $\mathcal{O}\left( 80\%\right) $ $%
n=2$ (from quasi-elastic scattering, $\nu_{\mu}n\rightarrow\mu^{-}p$,
neutral pion resonant production $\nu_{\mu}n\rightarrow\mu^{-}R^{+}%
\rightarrow\mu^{-}p\pi^{0}$, and coherent pion production $\nu_{\mu}\mathrm{%
Ar}\rightarrow \mu^{-}\pi^{+}\mathrm{Ar}$), $\mathcal{O}\left( 20\%\right) $ 
$n=3$ (resonant \ charged pion production $\nu_{\mu}p\rightarrow%
\mu^{-}R^{++}\rightarrow\mu^{-}p\pi^{+}$), and $\mathcal{O}\left( 1\%\right) 
$ $n\geq4$ (from multiparticle production processes referred to as
\textquotedblleft deep inelastic scattering\textquotedblright (DIS)). \ However final state interactions
(FSI) of hadrons produced in neutrino scattering with the argon nucleus can
subtract or add charged particles that emerge from within the nucleus. These multiplicities are further modified by the selection cuts.

\ We require tracks to register a minimum number of CP hits $(20)$ in the TPC
so that the Pandora MicroBooNE track reconstruction algorithms~\cite{Pandora
reference} operate optimally. Tracks with less than $20$ CP hits may fail to reconstruct due to inefficiencies in the reconstruction algorithms. This minimum CP hit condition corresponds to a
minimum range in liquid argon of 6 cm, and the requirement thus excludes
charged particles below a particle-type-dependent kinetic energy threshold
from entering our sample. This threshold ranges from $37$ MeV for a $%
\pi^{\pm }$ to $82$ MeV for a proton, and this measurement has no acceptance
for particles with kinetic energies below these thresholds. These thresholds are also dependent on the angle of the track with respect to the collection plane wires. 

The average MicroBooNE\ charged track reconstruction efficiency is $%
\left\langle \epsilon \right\rangle \approx 45\%$~\cite{DocDB5987-Pandora-PUB} at the $20$ hits threshold
used in this analysis. \ This relatively low
value, with implicit kinetic
energy thresholds, creates a common occurrence called \textquotedblleft
feed-down\textquotedblright\ wherein events produced with $n$ tracks at the
argon nucleus exit position are reconstructed with an observed multiplicity $%
n^{\prime }<n$. \ For example, $n=1$ is commonly observed because one of the
two tracks in a quasi-elastic event fails to be reconstructed. \ 

The underlying observed CPMD in MicroBooNE is expected to lie predominantly in
the $n=1-4$ range. \ The following summarizes qualitative expectations for
components of observed multiplicities from particular processes. \ These
components can include contributions from the primary neutrino-nucleon scatter
within the nucleus and secondary interactions of primary hadrons with the
remanent nucleus. \ Secondary charged particles are usually protons, which are
expected to be produced with kinetic energies that are usually too low for track reconstruction in this analysis. \ However, more energetic
forward-produced protons from the upper \textquotedblleft
tail\textquotedblright\ of this secondary kinetic energy distribution may make
it into our sample.

\begin{itemize}
\item Multiplicity $>3$, mainly predicted to be \textquotedblleft DIS events\textquotedblright%
\ in which at least three short tracks are reconstructed. \ \textquotedblleft
DIS\textquotedblright\ is the usual term for multiparticle final states not
identified with any particular resonance formation. \ Some contribution could
exist from multiplicity=3 resonant charged pion production accompanied by a proton from the high-energy tail of the final state interaction proton production distribution.

\item Multiplicity $=3$, mainly predicted to be $\mu^{-}p\pi^{+}$ events from $\Delta$
resonance production in which all three tracks are reconstructed.
\ \textquotedblleft Feed down\textquotedblright\ from higher multiplicity
would be small due to the tiny DIS cross section at MicroBooNE energies.
\ Some contribution could exist from multiplicity=2 QE scattering accompanied
by a proton from the high-energy tail of the final state interaction proton production distribution.

\item Multiplicity $=2$, mainly predicted to be QE $\mu^{-}p$ events in which the proton is
reconstructed, with a sub-leading contribution from ``feed down" of resonant
charged pion production events where one track fails to be reconstructed.
\ Multiplicity 2 could be augmented by high energy final state interaction protons.

\item Multiplicity $=1$, mainly predicted to be \textquotedblleft feed down\textquotedblright%
\ from QE $\mu^{-}p$ and $\mu^{-}p\pi^{0}$ events in which the proton is not
reconstructed, with contributions from other higher multiplicity topologies in
which more than one tracks fails to be reconstructed.
\end{itemize}

\begin{figure}[h]
\begin{minipage}{.5\textwidth}
\centering
\includegraphics[width=1.0\linewidth]{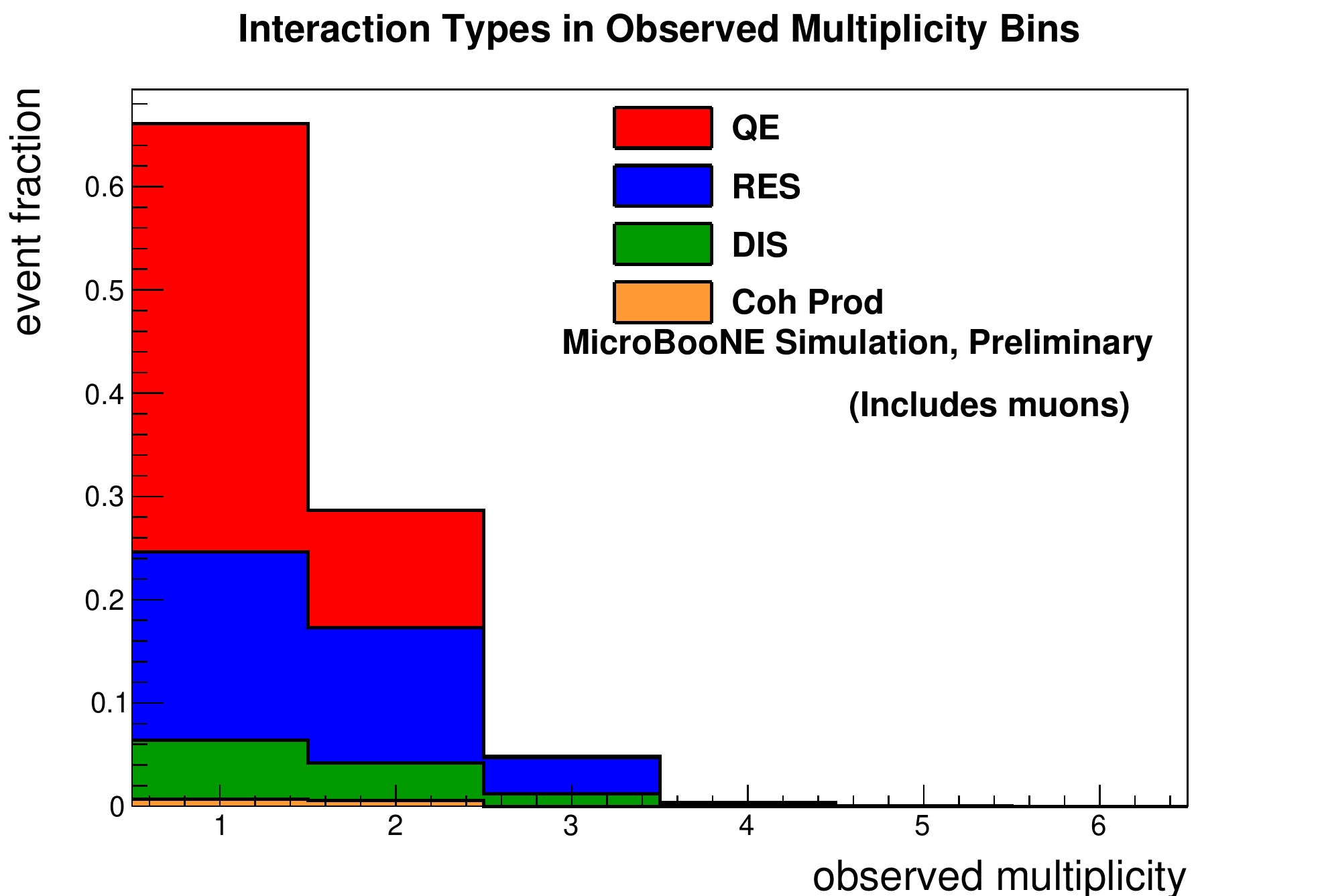}
\end{minipage}
\begin{minipage}{.5\textwidth}
\centering
\includegraphics[width=1.0\linewidth]{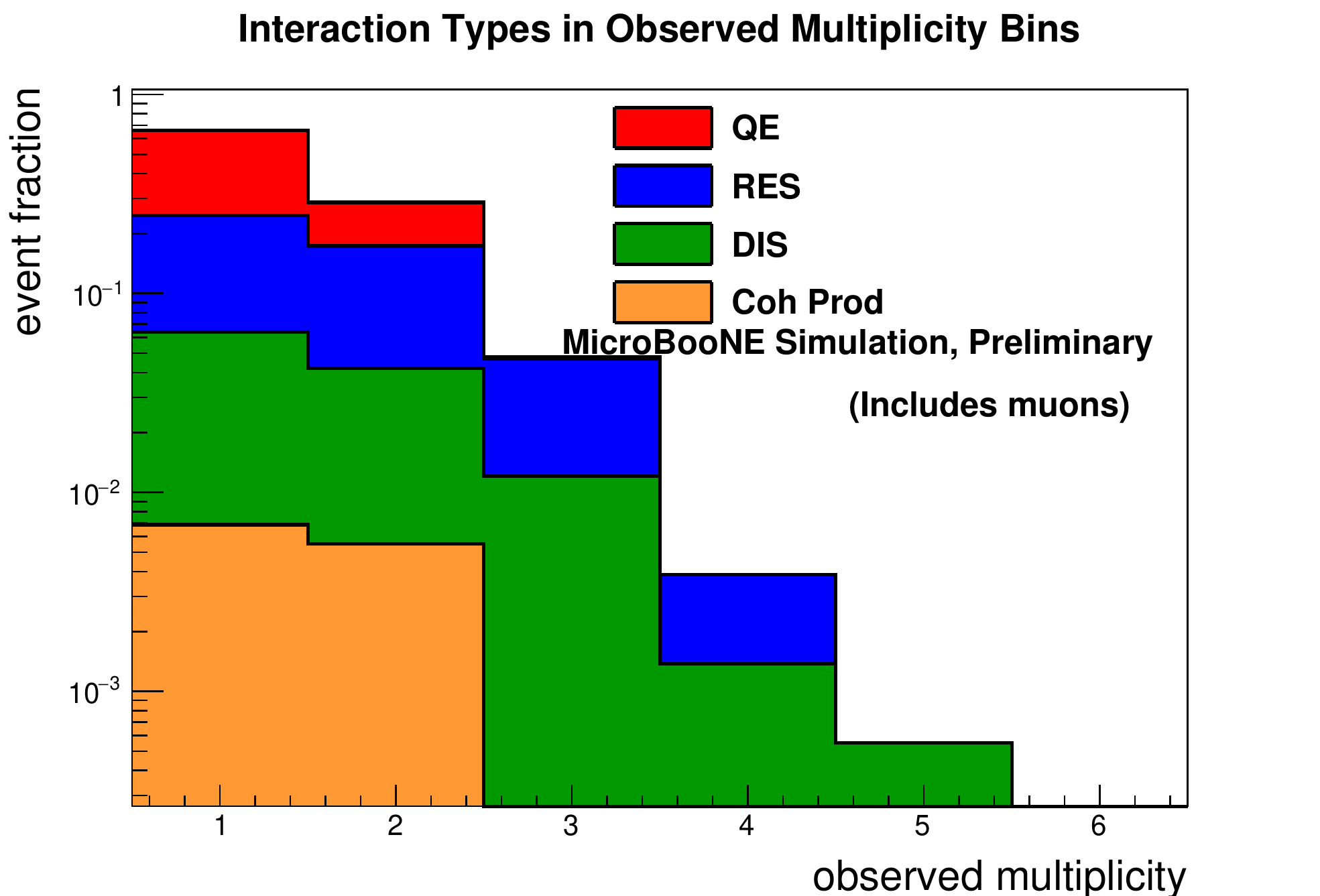}
\end{minipage}
\caption{Observed (stacked) multiplicity distributions for different
neutrino interaction types from our BNB-only default MC simulation in linear scale (left) and in log y scale (right).}
\label{img:inttype_MC_MCtrue}
\end{figure}

Figure \ref{img:inttype_MC_MCtrue} illustrates these expectations using our BNB-only MC simulation.
The $O_{2}/O_{1}$ ratio may be sensitive to the proton kinematics from QE
scattering. \ The $O_{3}/O_{2}$ ratio may provide sensitivity to the
value of the $\Delta$ resonance production cross section relative to the QE
cross section and to the pion kinematics from resonance decay and
propagation through the nucleus. \ Higher $n$ values of $O_{n}$ could test
for the DIS\ contribution and the presence of high energy tails in proton
production by final state interactions (FSI).

We note that our kinetic energy thresholds limit acceptance in such a way that protons produced in FSI may not significantly contribute to the observed CPMD. Furthermore, our analysis requires a forward-going long contained track as a muon candidate, which restricts the final state phase space. \ Our results should therefore not be compared to the low
energy proton multiplicity measurement reported by ArgoNeuT~\cite{ArgoNeuT
protons}. \ A future publication will be devoted to a measurement of proton
multiplicity in MicroBooNE over a larger phase space.

\section{Analysis Method and Results}

\label{sig_extraction}

\subsection{Cosmic ray Backgrounds in MicroBooNE}

The MicroBooNE\ detector lacks appreciable shielding from cosmic rays. \
Most events that pass trigger conditions during neutrino beam operations
(\textquotedblleft on-beam data\textquotedblright ) contain no neutrino
interactions, and triggered events with a neutrino interaction typically
have the products of several cosmic rays in the event readout window
contributing to the detector response along with the products of the neutrino
collision. \ A large sample of events recorded under identical conditions as
the on-beam data, except for the coincidence requirement with the beam,
(\textquotedblleft off-beam-data\textquotedblright ) has been recorded for
use in characterizing CR backgrounds. \ A straightforward on-beam minus
off-beam background subtraction is, however, difficult, as the off-beam data
does not reproduce all correlated detector effects associated with on-beam
events containing a neutrino interaction with several overlaid cosmic rays.
\ The situation is particularly complicated with observed multiplicity$=$1
neutrino interaction events, which share a common topology with the abundant single muon CR background MC simulations of the CR flux using the CORSIKA package provide useful guidance; however, the ability of
these simulations to describe the very rare CR topologies that closely match
neutrino interactions is not yet well understood.

For these reasons, this analysis employs a method to separate neutrino
interaction candidates from CR backgrounds that is driven by the data
itself. \ The separation rests on the observation that a neutrino $\nu _{\mu
}$ CC interaction produces a final state $\mu ^{-}$ that slows down as it
moves away from its production point at the neutrino interaction vertex due
to ionization energy loss in the liquid argon. \ As it slows down, its rate
of restricted energy loss, $dE/dx_{R}$, increases, and deviations from a
linear trajectory due to multiple Coulomb scattering (MCS) become more
pronounced. \ A CR muon track can produce an apparent neutrino interaction
vertex if it comes to rest in the detector, but the CR track will exhibit
large $dE/dx_{R}$ and MCS effects in the vicinity of this vertex. \
Furthermore, the vast majority of $\nu _{\mu }$ CC muons that satisfy the $75$
cm length requirement travel in the neutrino beam direction
(\textquotedblleft upstream\textquotedblright\ to \textquotedblleft
downstream\textquotedblright ), whereas CR muons move upstream or downstream
with equal probability.

\subsection{Data-Driven Signal+Background Model}

On-beam data consists of a mixture of neutrino interaction and CR background
events. \ For each observed track multiplicity, we divide the data sample into
four data categories, denoted as \textquotedblleft$\nu\nu$\textquotedblright, \textquotedblleft$CR\nu
$\textquotedblright, \textquotedblleft$\nu CR$\textquotedblright, and \textquotedblleft$CRCR$\textquotedblright\ according
to the outcome of the PH and MCS tests performed on the longest track in the
event. \ These samples contain numbers of events equal to $N_{\nu\nu}$,
$N_{CR\nu}$, $N_{\nu CR},$ and $N_{CRCR}$, respectively. \ The
\textquotedblleft$\nu$\textquotedblright\ designation indicates that the $PH$
or $MCS$ test categorizes the long track as being more likely a $\nu_{\mu}$ CC
event candidate muon, while the \textquotedblleft$CR$\textquotedblright%
\ designation corresponds to a long track candidate falling into the more
likely CR class according to the $PH$ or $MCS$ test. \ The number of events in
each category, for each multiplicity, is then modelled by the following:
\begin{align}
\hat{N}_{\nu\nu} &  =P\left(  MCS|PH\right)  P\left(  PH\right)  \hat{N}_{\nu
}\label{model 1}\\
&  +\left(  1-Q\left(  PH\right)  -Q\left(  MCS\right)  +Q\left(
MCS|PH\right)  Q\left(  PH\right)  \right)  \hat{N}_{CR},\nonumber\\
\hat{N}_{CR\nu} &  =\left(  1-P\left(  MCS|PH\right)  \right)  P\left(
PH\right)  \hat{N}_{\nu}\label{model 2}\\
&  +\left(  Q\left(  MCS\right)  -Q\left(  MCS|PH\right)  Q\left(  PH\right)
\right)  \hat{N}_{CR},\nonumber\\
\hat{N}_{\nu CR} &  =\left(  P\left(  MCS\right)  -P\left(  MCS|PH\right)
P\left(  PH\right)  \right)  \hat{N}_{\nu}\label{model 3}\\
&  +\left(  1-Q\left(  MCS|PH\right)  \right)  Q\left(  PH\right)  \hat
{N}_{CR},\nonumber\\
\hat{N}_{CRCR} &  =\left(  1-P\left(  PH\right)  -P\left(  MCS\right)
+P\left(  MCS|PH\right)  P\left(  PH\right)  \right)  \hat{N}_{\nu
}\label{model 4}\\
&  +Q\left(  MCS|PH\right)  Q\left(  PH\right)  \hat{N}_{CR},\nonumber
\end{align}
where $\hat{N}_{\nu\nu}$ corresponds to the number of event in the sample that
passes both PH and MCS tests and $\hat{N}_{CRCR}$ corresponds to the number of
events in the sample that fails both PH and MCS tests. These are expected to
be the samples with enriched neutrino and CR content, respectively, $\hat{N}_{CR\nu}$ and $\hat{N}_{\nu CR}$ corresponds to the number of events
in the sample that pass one test and fails the other. These are the samples
of mixed purity. \ The quantities $\hat{N}_{\nu}$ and $\hat{N}_{CR}$ are the
to-be-fitted number of neutrino and CR events, respectively, in the sample. \ 

The conditional probability $P\left(  MCS|PH\right)  $ denotes the fraction of
time that an event passes the $MCS$ condition after it has passed the $PH$
condition. \ As the $MCS$ and $PH$ conditions result from different physical
processes (muon-nucleus and muon-electron scattering, respectively) and the
$MCS$ and $PH$ test are formed primarily from different quantities (time and
charge, respectively), the PH and MCS tests are nearly independent, $P\left(
MCS|PH\right)  \approx$ $P\left(  MCS\right)  $. \ In the analysis we find
evidence for weak correlations between the tests, and use of the conditional
probability allows for this to be taken into account.

Off-beam data, which contains no neutrino content, is divided into the same
categories, with event counts $N_{\nu\nu}^{\prime}$, $N_{CR\nu}^{\prime}$,
$N_{\nu CR}^{\prime},$ and $N_{CRCR}^{\prime}$, and modeled as%
\begin{align}
\hat{N}_{\nu\nu}^{\prime} &  =\left(  1-Q\left(  PH\right)  -Q\left(
MCS\right)  +Q\left(  MCS|PH\right)  Q\left(  PH\right)  \right)  \hat{N}%
_{CR}^{\prime},\label{model 5}\\
\hat{N}_{CR\nu}^{\prime} &  =\left(  Q\left(  MCS\right)  -Q\left(
MCS|PH\right)  Q\left(  PH\right)  \right)  \hat{N}_{CR}^{\prime
},\label{model 6}\\
\hat{N}_{\nu CR}^{\prime} &  =\left(  1-Q\left(  MCS|PH\right)  \right)
Q\left(  PH\right)  \hat{N}_{CR}^{\prime},\label{model 7}\\
\hat{N}_{CRCR}^{\prime} &  =Q\left(  MCS|PH\right)  Q\left(  PH\right)
\hat{N}_{CR}^{\prime}.\label{model 8}%
\end{align}
where $\hat{N}_{\nu\nu}^{\prime}$ and $\hat{N}_{CRCR}^{\prime}$ are expected
to be enriched samples containing muons characteristic of neutrino
interactions and cosmic rays, respectively, and $\hat{N}_{CR\nu}^{\prime}$ and
$\hat{N}_{\nu CR}^{\prime}$ are samples of mixed purity; and \ $\hat{N}%
_{CR}^{\prime}$ is the to-be-fitted CR content of the sample (in practice the
number of events in the sample).

Two parameters, $\alpha _{\nu }$and  $\alpha _{\text{CR}}$ describe the
conditional probabilities $P\left( MCS|PH\right) $ and $Q\left(
MCS|PH\right) $ via the parameterizations given below$.$ These are calculated
from the BNB-only MC simulation and off-beam data, respectively. \
These are parameterized as%
\begin{align}
P\left( MCS|PH\right) & =\frac{\alpha _{\nu }P\left( MCS\right) }{1+\left(
\alpha _{\nu }-1\right) P\left( MCS\right) }, \\
Q\left( MCS|PH\right) & =\frac{\alpha _{\text{CR}}Q\left( MCS\right) }{%
1+\left( \alpha _{\text{CR}}-1\right) Q\left( MCS\right) }.
\end{align} 

Our algorithm uses the eight categories of events (Eqs. \ref{model 1}%
-\ref{model 8}) in on-beam and off-beam data to fit for the neutrino content
in each observed multiplicity bin.

\subsection{Fitting Procedure}

We construct a likelihood function based on the probability distribution for
partitioning events into one of four categories, a multinomial distribution.
\ The multinomial probability of observing $n_{i}$ events in bin $i$, with $%
i=1,2,3,4$, with the probability of a single event landing in bin $i$ equal
to $r_{i}$ is

\begin{equation}
M\left( n_{1},n_{2},n_{3},n_{4};r_{1},r_{2},r_{3},r_{4}\right) =\frac{\left(
n_{1}+n_{2}+n_{3}+n_{4}\right) !}{n_{1}!n_{2}!n_{3}!n_{4}!}%
r_{1}^{n_{1}}r_{2}^{n_{2}}r_{3}^{n3}r_{4}^{n_{4}}.
\end{equation}
The $n_{i}$ will be the observed number of events in each multiplicity bin, and the $%
r_{i} $ will be functions of the model fit parameters:

\begin{align}
\text{on-beam}\text{: } & M\left( N_{\nu\nu},N_{CR\nu},N_{\nu CR},N_{CRCR};%
\frac{\hat{N}_{\nu\nu}}{\hat{N}},\frac{\hat{N}_{CR\nu}}{\hat{N}},\frac{\hat{N%
}_{\nu CR}}{\hat{N}},\frac{\hat{N}_{CRCR}}{\hat{N}}\right) , \\
\text{off-beam}\text{: } & M\left( N_{\nu\nu}^{\prime},N_{CR\nu}^{\prime
},N_{\nu CR}^{\prime},N_{CRCR}^{\prime};\frac{\hat{N}_{\nu\nu}^{\prime}}{%
\hat{N}^{\prime}},\frac{\hat{N}_{CR\nu}^{\prime}}{\hat{N}^{\prime}},\frac{%
\hat{N}_{\nu CR}^{\prime}}{\hat{N}^{\prime}},\frac{\hat{N}_{CRCR}^{\prime}}{\hat{N}%
^{\prime}}\right) ,
\end{align}
with%
\begin{align}
\hat{N} & =\hat{N}_{\nu\nu}+\hat{N}_{CR\nu}+\hat{N}_{\nu CR}+\hat{N}_{CRCR},
\\
\hat{N}^{\prime} & =\hat{N}_{\nu\nu}^{\prime}+\hat{N}_{CR\nu}^{\prime}+\hat{N%
}_{\nu CR}^{\prime}+\hat{N}_{CRCR}^{\prime}.
\end{align}

The likelihood also incorporates the Poisson statistics of observing $%
n_{1}+n_{2}+n_{3}+n_{4}$ in both the on-beam and off-beam data:%
\begin{align}
\text{on-beam}\text{: } & \frac{\hat{N}^{N}}{N!}e^{-\hat{N}}, \\
\text{off-beam}\text{: } & \frac{\hat{N}^{\prime N^{\prime}}}{N^{\prime}!}%
e^{-\hat{N}^{\prime}},
\end{align}
with%
\begin{align}
N & =N_{\nu\nu}+N_{CR\nu}+N_{\nu CR}+N_{CRCR}, \\
N^{\prime} & =N_{\nu\nu}^{\prime}+N_{CR\nu}^{\prime}+N_{\nu CR}^{\prime
}+N_{CRCR}^{\prime}.
\end{align}

The final likelihood function is 
\begin{align}
L_{TOT} & =M\left( N_{\nu\nu},N_{CR\nu},N_{\nu CR},N_{CRCR};\frac{\hat {N}%
_{\nu\nu}}{\hat{N}},\frac{\hat{N}_{CR\nu}}{\hat{N}},\frac{\hat{N}_{\nu CR}}{%
\hat{N}},\frac{\hat{N}_{CRCR}}{\hat{N}}\right) \\
& \times M\left( N_{\nu\nu}^{\prime},N_{CR\nu}^{\prime},N_{\nu CR}^{\prime
},N_{CRCR}^{\prime};\frac{\hat{N}_{\nu\nu}^{\prime}}{\hat{N}^{\prime}},\frac{%
\hat{N}_{CR\nu}^{\prime}}{\hat{N}^{\prime}},\frac{\hat{N}_{\nu CR}^{\prime}}{%
\hat{N}^{\prime}},\frac{\hat{N}_{CRCR}^{\prime}}{\hat{N}^{\prime}}\right)  \notag
\\
& \times\frac{\hat{N}^{N}}{N!}e^{-\hat{N}}\times\frac{\hat{N}^{\prime
N^{\prime}}}{N^{\prime}!}e^{-\hat{N}^{\prime}}.  \notag
\end{align}
The model fit parameters and their statistical uncertainties are estimated via
the maximum likelihood method, implemented by minimizing the
negative-log-likelihood

\begin{equation}
\mathcal{L}_{TOT}=-\ln L_{TOT}\text{,}
\end{equation}
using the MIGRAD minimization in the standard MINUIT~\cite{MINIUT} package in ROOT~\cite{ROOT}.

\begin{figure}[tbp]
\centering
\includegraphics[width=0.9\textwidth]{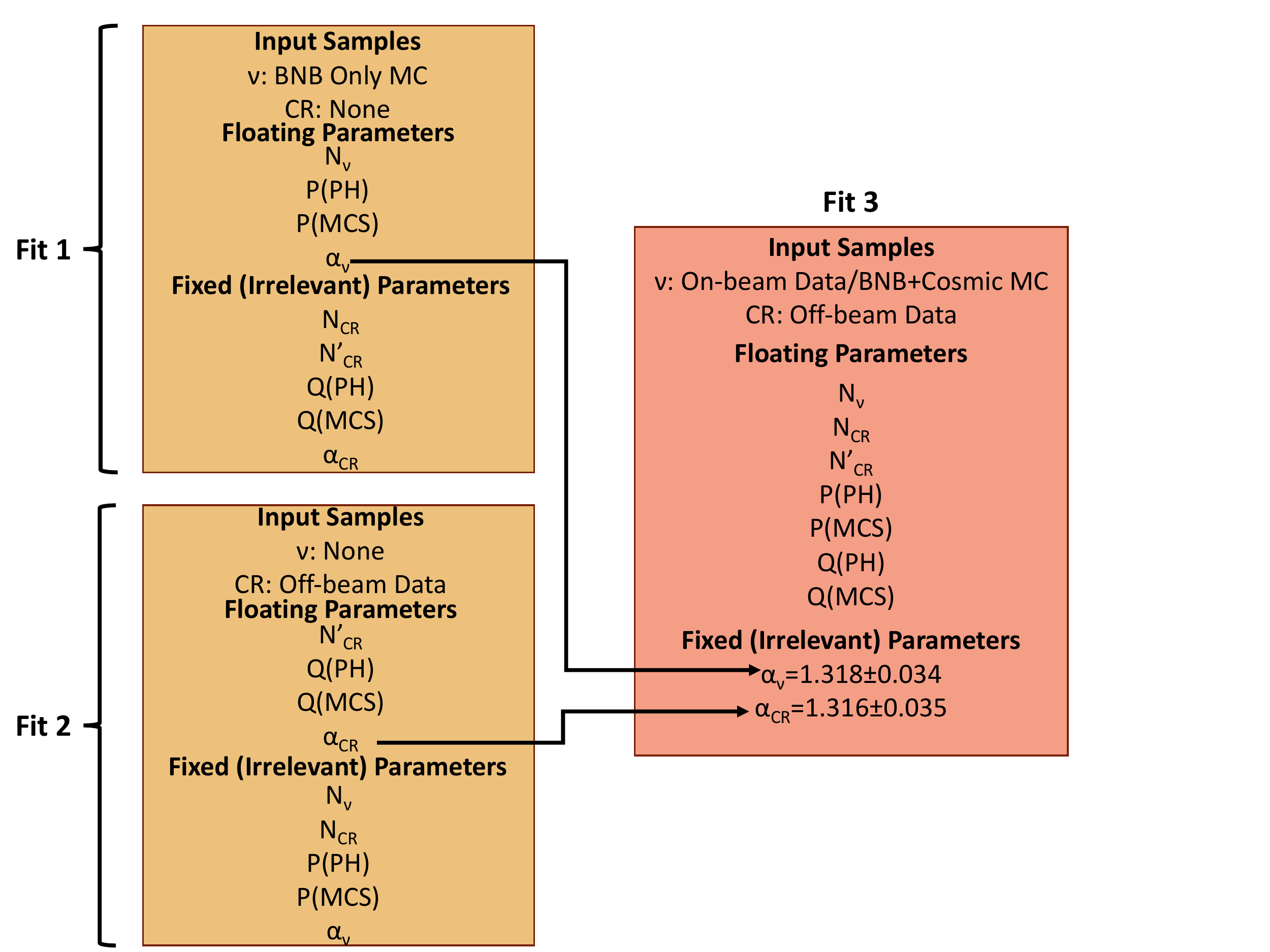} 
\caption{Fitting process diagram.  Fixed and floating parameters and input samples are listed for each fit. }
\label{img:fit_workflow}
\end{figure}

The fitting procedure can be used to obtain estimates for $\hat{N}_{\nu }$, $%
\hat{N}_{CR}$, $\hat{N}_{CR}^{\prime }$, $P\left( PH\right) $, $P\left(
MCS\right) $, $Q\left( PH\right) $, and $Q\left( MCS\right) $ for each
multiplicity. \ When the probability parameters are consistent between
multiplicities, we use all multiplicities together in their determination
for improved statistical precision and vary the three parameters $\hat{N}%
_{\nu }$, $\hat{N}_{CR}$, and $\hat{N}_{CR}^{\prime }$ for each multiplicity.

\ In total, we have eight equations  (Eqs. \ref{model 1}$-$\ref{model 8}), where first four come from on-beam data and the other four come from off-beam data. We have nine parameters $\hat{N}_{\nu }$, $\hat{N}_{CR}$, $\hat{N}_{CR}^{\prime }$, $P(PH)$, $Q(PH)$, $P(MCS)$, $Q(MCS)$, $\alpha_\nu$ and $\alpha_{CR}$. After proving that the off-beam data and CR-only MC are consistent, the parameters $\alpha_\nu$ and $\alpha_{CR}$ were obtained from BNB-only MC simulation and off-beam data samples, respectively, and were kept fixed afterwards. The maximum likelihood fit was performed to obtain the values of the rest of seven parameters from simulation (BNB+Cosmic default MC $\&$ off-beam data) and data (on-beam $\&$ off-beam data) samples. \ Figure \ref{img:fit_workflow} presents a schematic diagram for the fitting process and also lists the fixed and floating parameters for each fit.

\subsection{Results on Simulation}

A maximum likelihood fit was performed on all three different simulation
samples. This fit was performed to extract the values of seven parameters $%
\hat{N}_{\nu }$, $\hat{N}_{CR}$, $\hat{N}_{CR}^{\prime }$, $P(PH)$, $Q(PH)$, $%
P(MCS)$, and $Q(MCS)$). As expected, the $PH$ and $MCS$ probabilities show
no statistically significant difference between the three GENIE\ models
considered. \ Table \ref{tab:fit_values_BNB+Cosmic_data} lists the fit
values obtained from the fit for the above-mentioned parameters in the
BNB+Cosmic default MC and the off-beam data.

\begin{table}[ptb]
\caption{Fit parameter results and corresponding errors for the BNB+Cosmic default MC simulation and off-beam data samples as well as the on-beam and off-beam data samples.  The same off-beam data sample was used in both fits. \ All errors are from the fit and are purely
statistical.}
\label{tab:fit_values_BNB+Cosmic_data}
\begin{center}
\begin{tabular}{c|c|c}
& \multicolumn{2}{c}{\textbf{Fit Results}} \\ 
\textbf{Parameters} & \textbf{BNB+Cosmic MC} & \textbf{MicroBooNE Data} \\ \hline
$\hat{N}_{\nu}$ & 3602$\pm$154 & 1056$\pm$169 \\ 
$\hat{N}_{CR}$ & 607$\pm$144 & 865$\pm$169 \\ 
$\hat{N}_{CR}^{\prime}$ & 5267$\pm$73 & 5267$\pm$73 \\ 
$P(PH)$ & 0.859$\pm$0.017 & 0.784$\pm$0.052 \\ 
$P(MCS)$ & 0.775$\pm$0.012 & 0.732$\pm$ 0.038 \\ 
$Q(PH)$ & 0.554$\pm$0.007 & 0.554$\pm$0.007 \\ 
$Q(MCS)$ & 0.544$\pm$0.007 & 0.544$\pm$ 0.007 \\ 

\end{tabular}
\end{center}
\end{table}

\subsection{Closure Test Results}

\begin{figure}[h]
\begin{minipage}{.5\textwidth}
\centering
\includegraphics[width=1.0\linewidth]{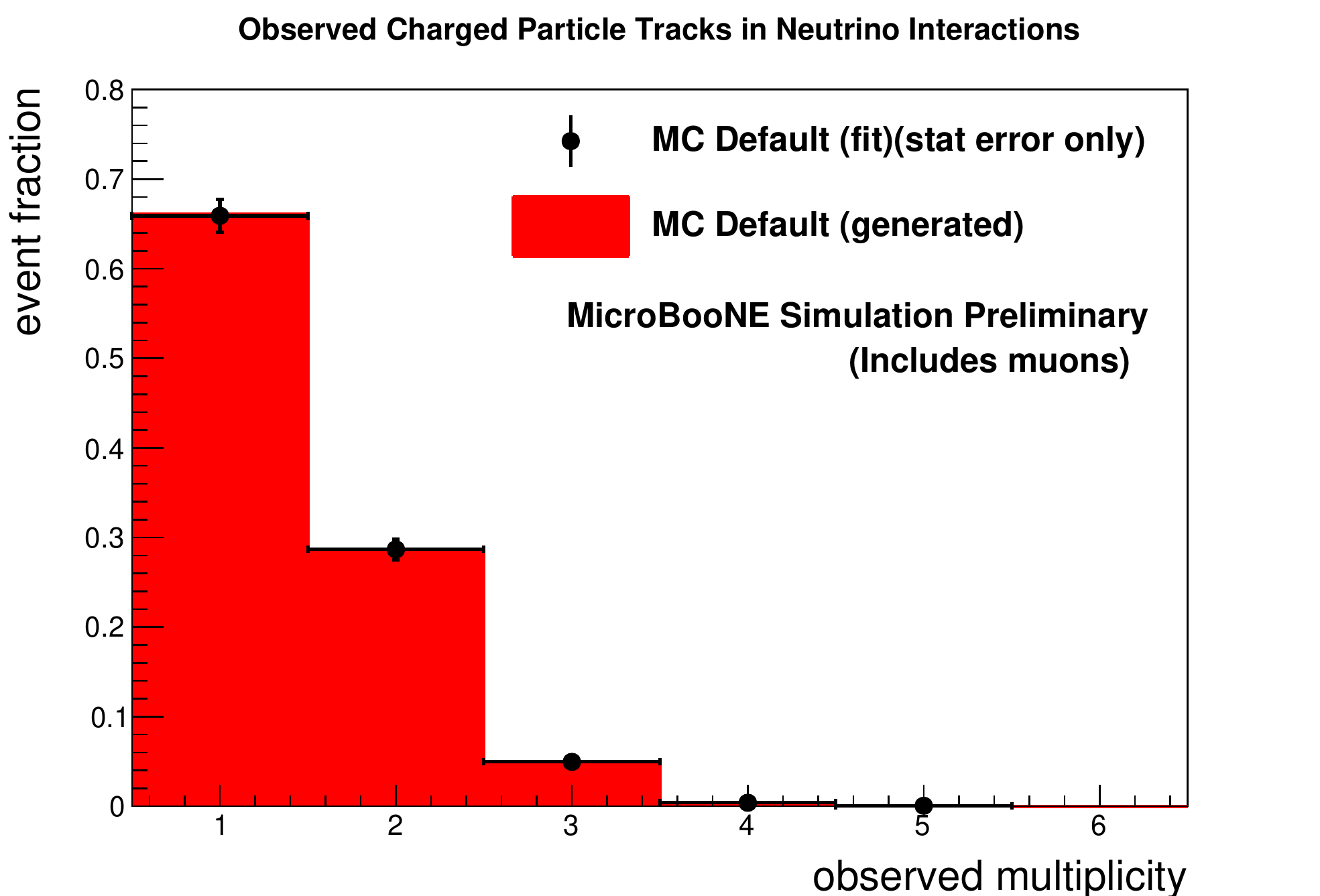}
\end{minipage}
\begin{minipage}{.5\textwidth}
\centering
\includegraphics[width=1.0\linewidth]{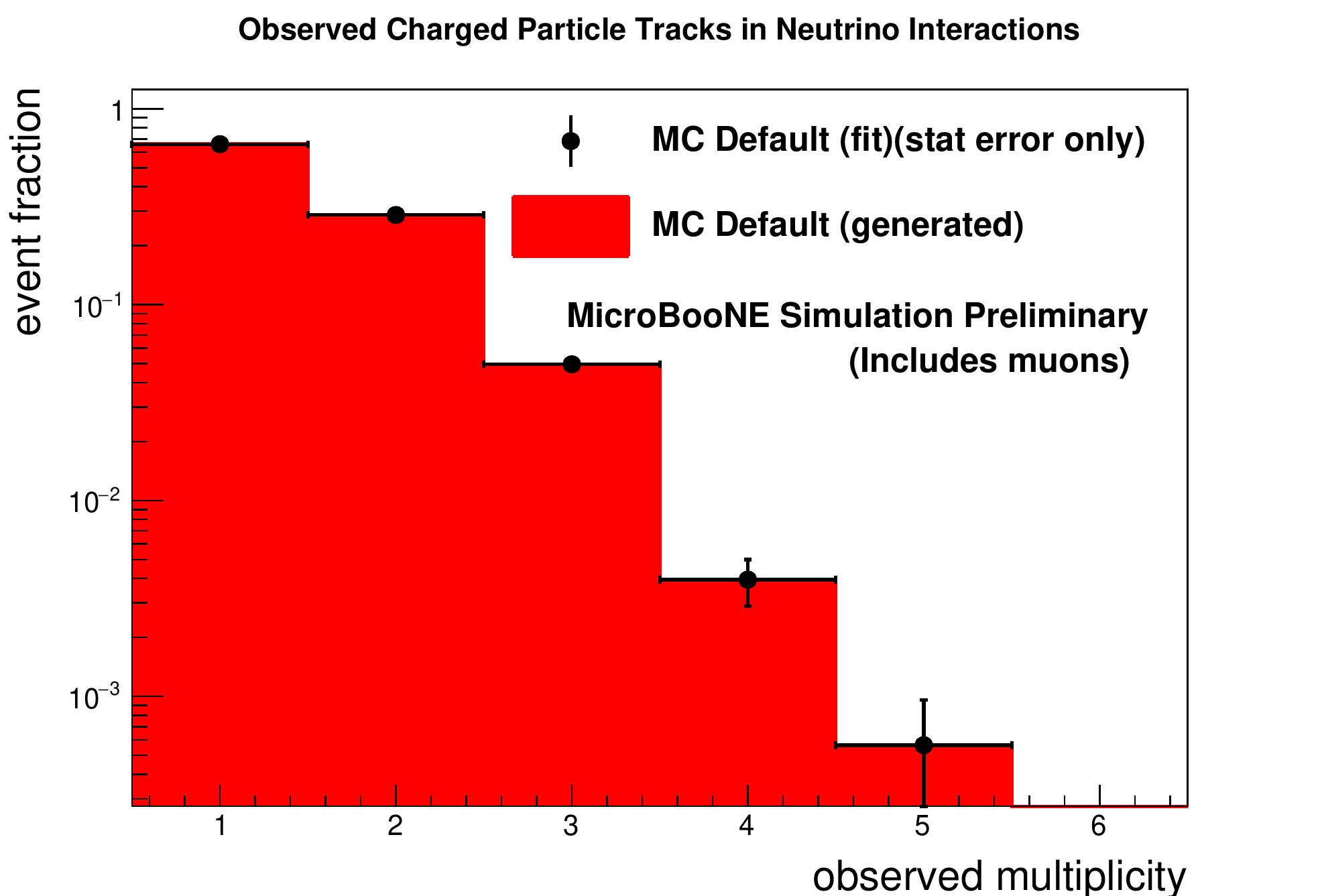}
\end{minipage}
\caption{Bin-by-bin fitted, area normalized, and  CR-background-subtracted
observed neutrino multiplicity distribution from the BNB+Cosmic default MC sample in linear scale (left) and in log y scale (right).}
\label{img:mult_MC_MCtrue}
\end{figure}

The number of neutrino events in the simulated data samples were extracted
and compared to the known number from the event generation. Table \ref%
{tab:mult_MC+cosmic} and Figure \ref{img:mult_MC_MCtrue} summarize this
comparison. We find that fit results agree within statistics with the known
inputs, indicating a lack of bias in our signal estimation technique. \ We
have also verified that our method is insensitive to the
signal-to-background ratio of the sample over a range corresponding to $%
0.2-5.0\ $times that estimated in the data.

\begin{table}[ptb]
\caption{Fitted and true number of neutrino events for the BNB+Cosmic default MC
sample for different multiplicity bins. The last column shows good agreement
between the fit results and true content for different bins.}
\label{tab:mult_MC+cosmic}
\begin{center}
\begin{tabular}{c|c|c|c}
\textbf{Multiplicities} & \textbf{Fit $N_\nu$} & \textbf{True $N_\nu$} &\textbf{True-Fit $\chi^{2}$/ndf} \\ \hline
1 & 2340$\pm$65 & 2405 & 1.0 \\ 
2 & 1018$\pm$41 & 1043 & 0.4 \\ 
3 & 176$\pm$13 & 175 & 0.0 \\ 
4 & 14$\pm$3 & 14 & 0.0 \\ 
5 & 2$\pm$1 & 2 & 0.0 \\ 
\end{tabular}
\end{center}
\end{table}

\subsection{Statistical and Systematic Uncertainty Estimates}

\subsubsection{Statistical Uncertainties}

\begin{figure}[tp]
\begin{adjustwidth}{0cm}{0cm}
\centering
\subfloat[][MC statistical uncertainty impact; linear scale.]
   {\includegraphics[width=.4\textwidth]{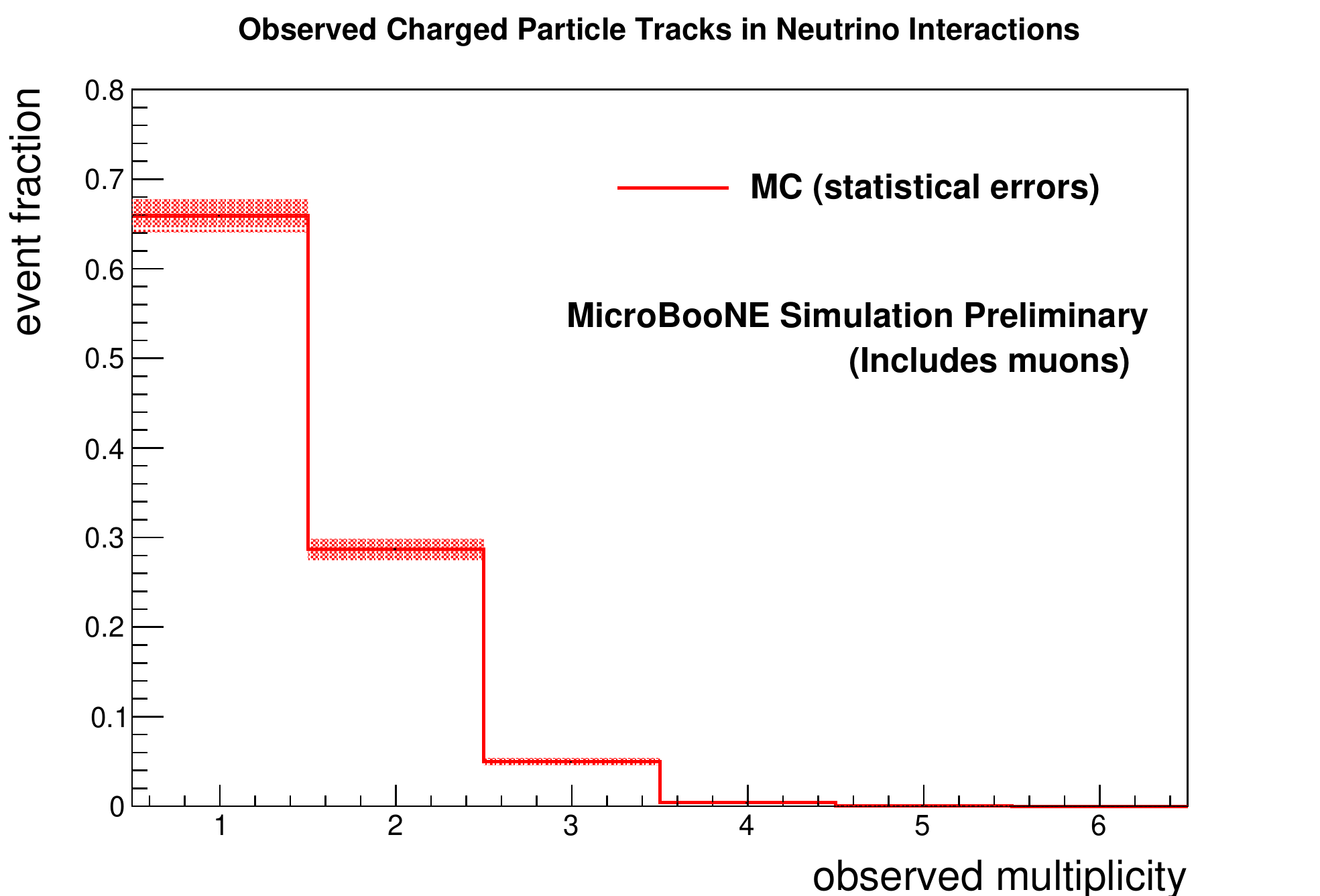}
   \label{img:stat_uncer(a)}} \quad
\subfloat[][MC statistical uncertainty impact; log y scale.]
   {\includegraphics[width=.4\textwidth]{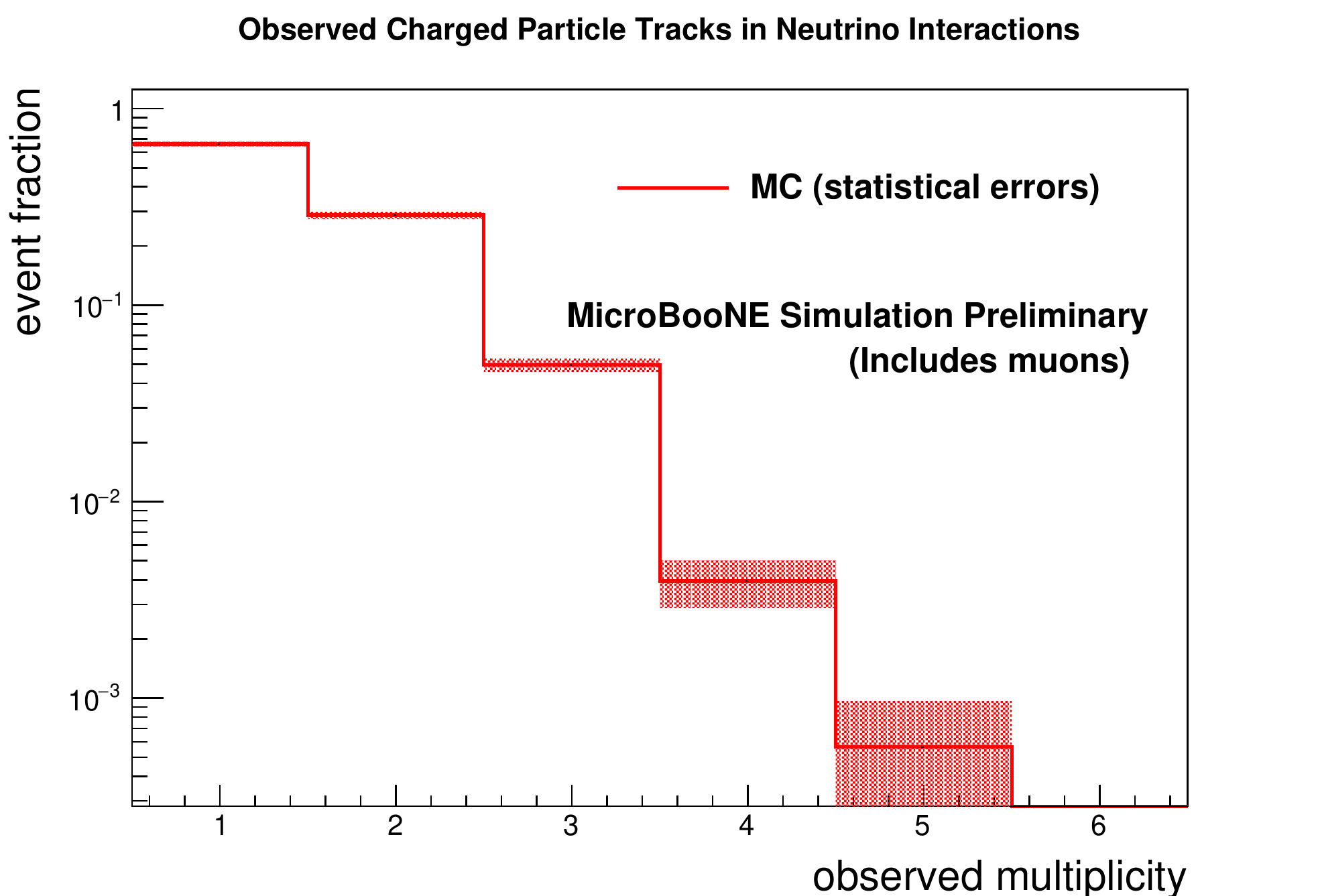}
   \label{img:stat_uncer(b)}} \quad
   
\subfloat[][Short track efficiency uncertainty impact; linear scale.]
   {\includegraphics[width=.4\textwidth]{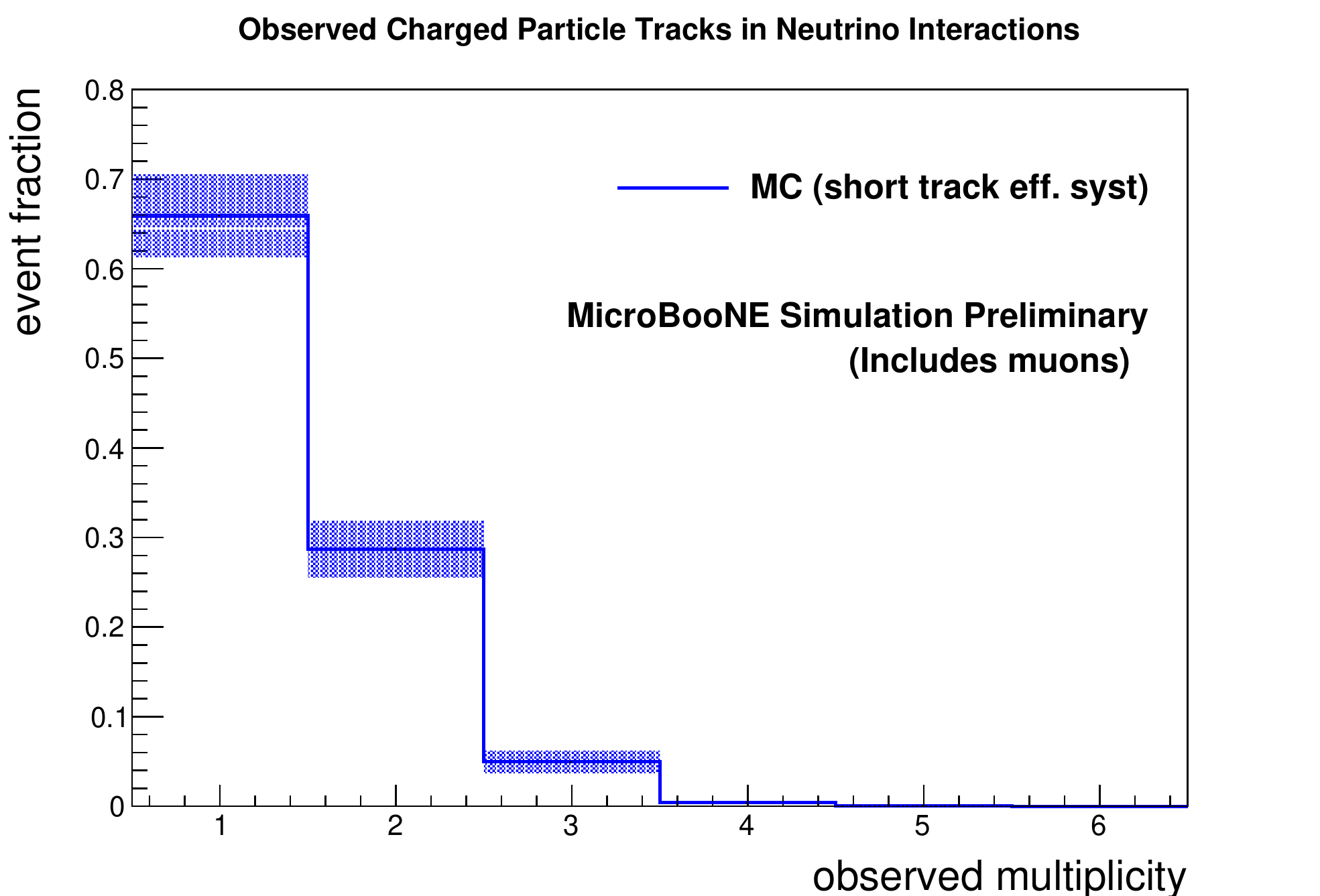}
   \label{img:eff_short_uncer(a)}} \quad
\subfloat[][Short track efficiency uncertainty impact; log y scale.]
   {\includegraphics[width=.4\textwidth]{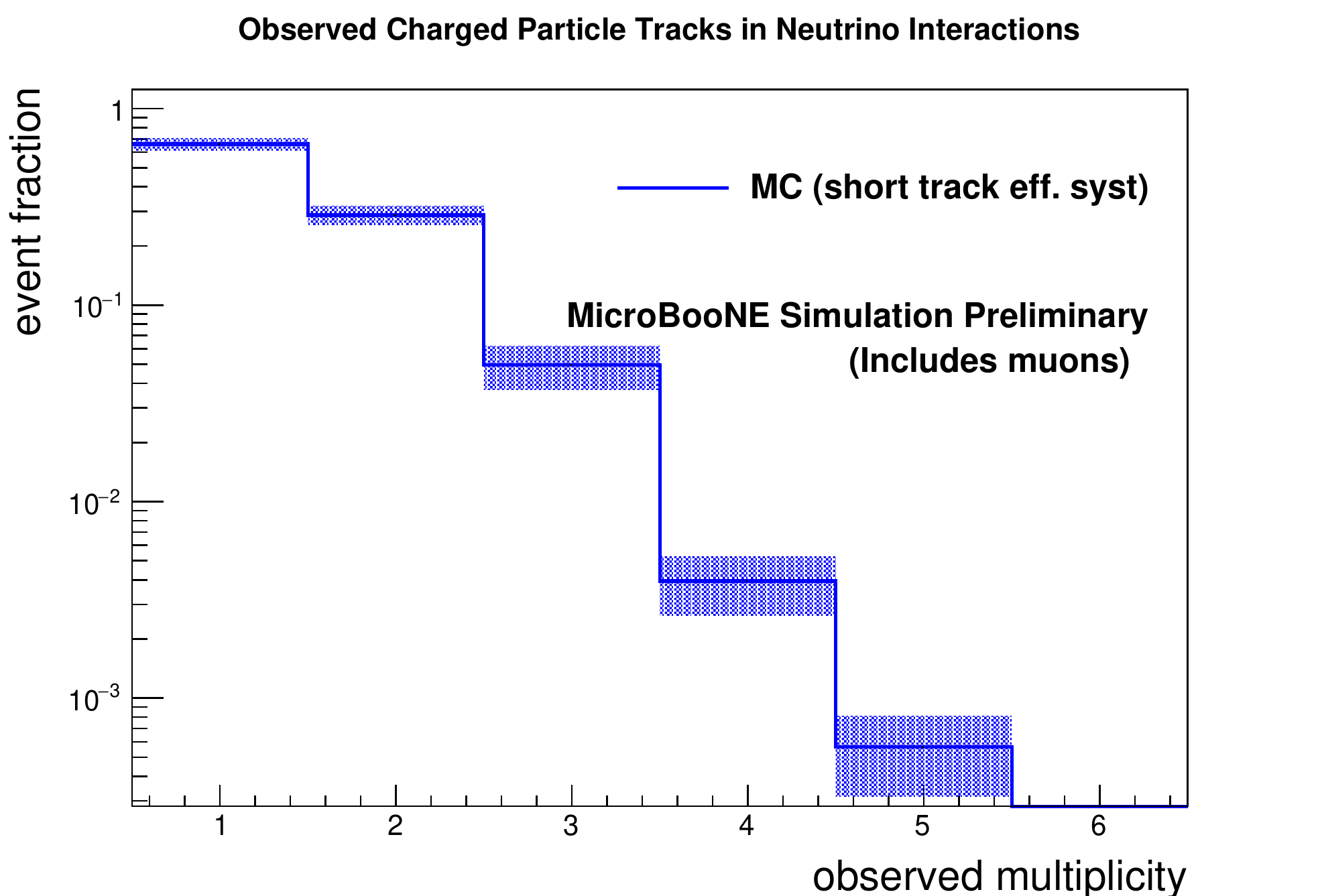}
   \label{img:eff_short_uncer(b)}} \quad  

\subfloat[][Long track efficiency uncertainty impact; linear scale.]
   {\includegraphics[width=.4\textwidth]{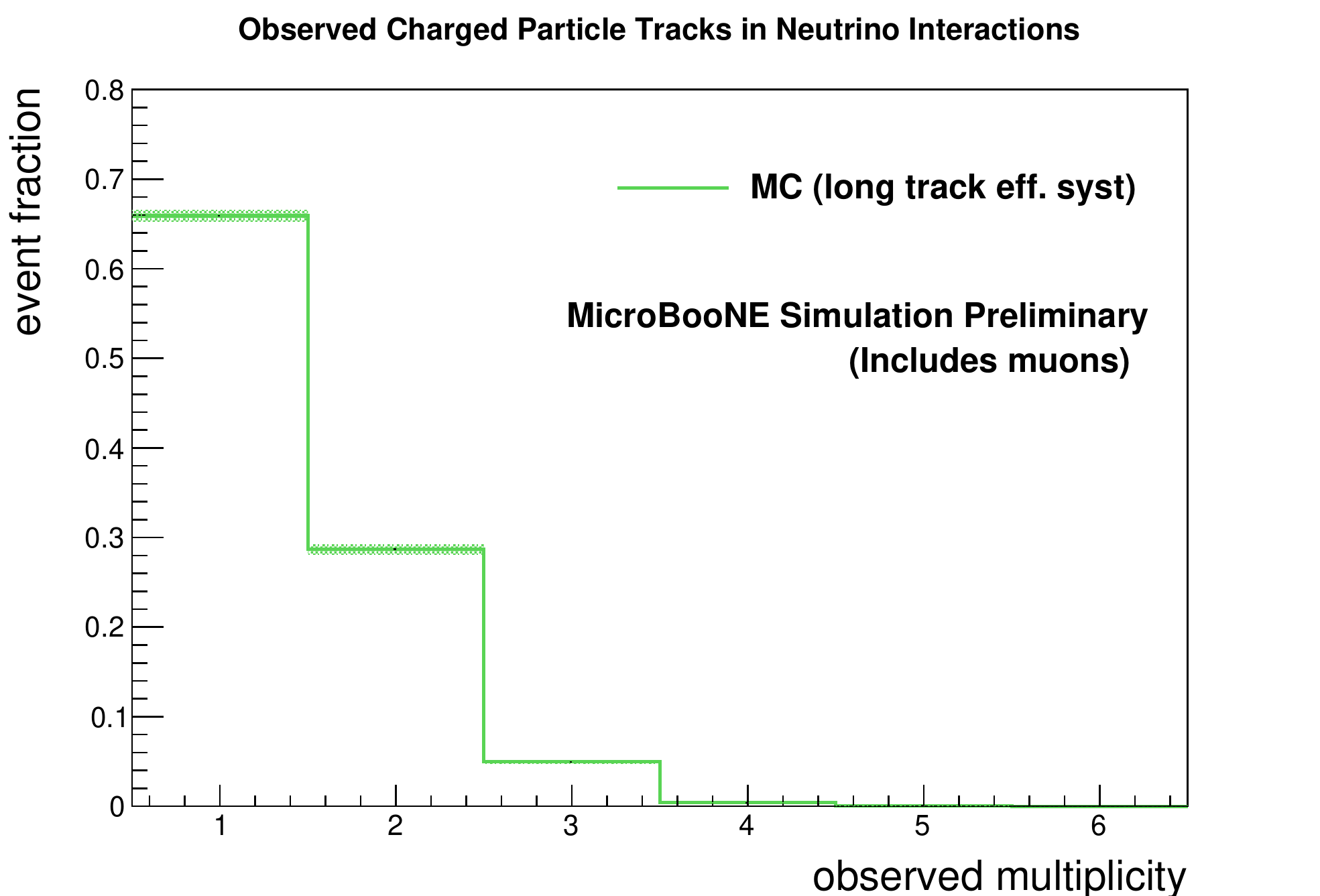}
   \label{img:eff_long_uncer(a)}} \quad
\subfloat[][Long track efficiency uncertainty impact; log y scale.]
   {\includegraphics[width=.4\textwidth]{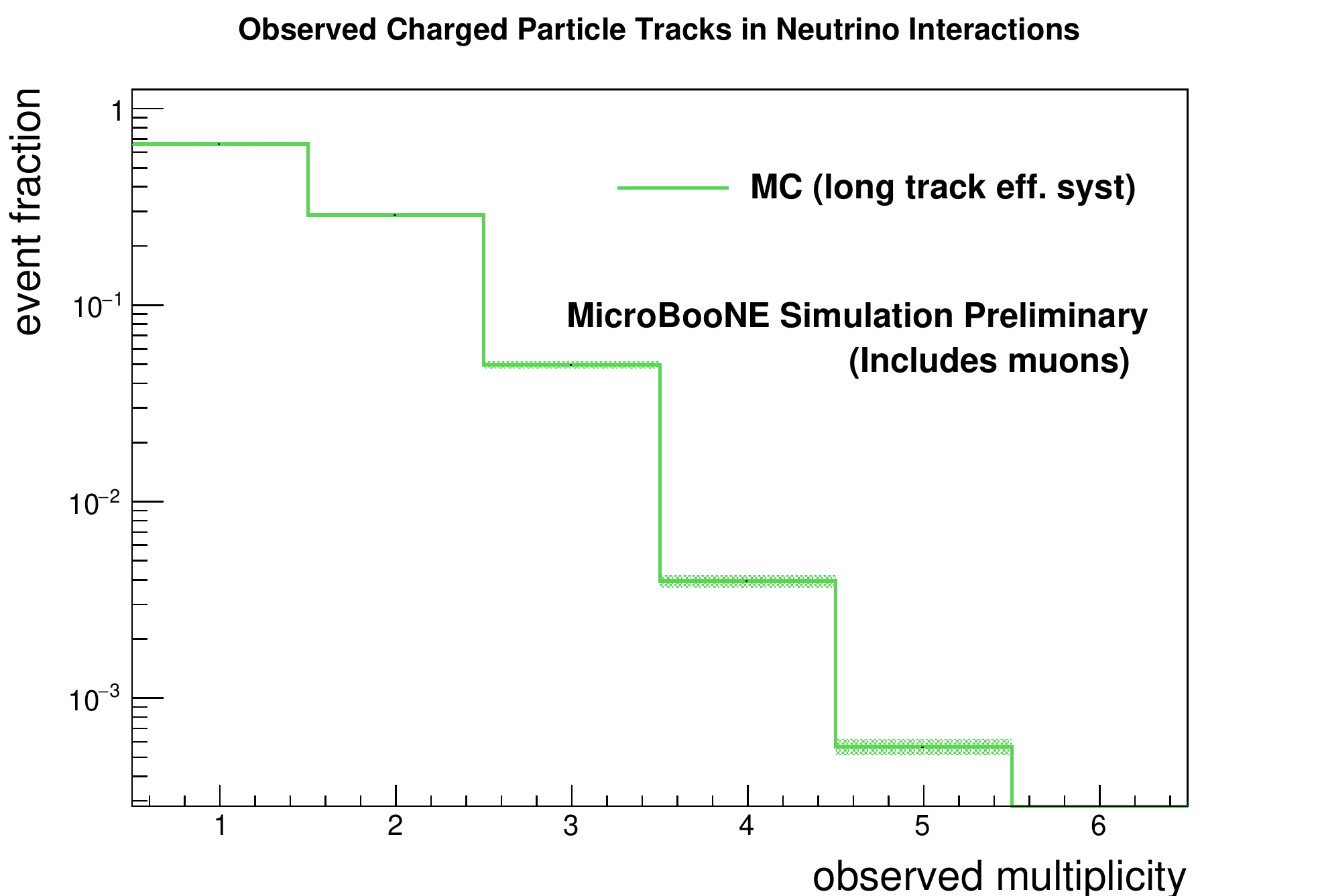}
   \label{img:eff_long_uncer(b)}} \quad 
   
\caption{Systematic uncertainty contributions to observed CPMD due to MC statistics (top row), short track efficiency (middle row), and long track efficiency (bottom row). The width of the line on the histogram indicates the uncertainty band.}
\label{img:sys_effects1}
\end{adjustwidth}
\end{figure}

Statistical uncertainties are returned from the MINUIT package used in our
fitting for both data and MC samples. These uncertainties include
contributions from the CR background in our fitting procedure. \ Both data
and MC statistics contribute substantially to the overall uncertainties in
our data, as shown in Figures \ref{img:stat_uncer(a)} and \ref%
{img:stat_uncer(b)}. These will be reduced in the future by employing the
full MicroBooNE data set from its first run and by generating larger MC
samples.

\begin{figure}[tp]
\begin{adjustwidth}{-0cm}{-0cm}
\centering

\subfloat[][Background model uncertainty impact; linear scale.]
   {\includegraphics[width=.4\textwidth]{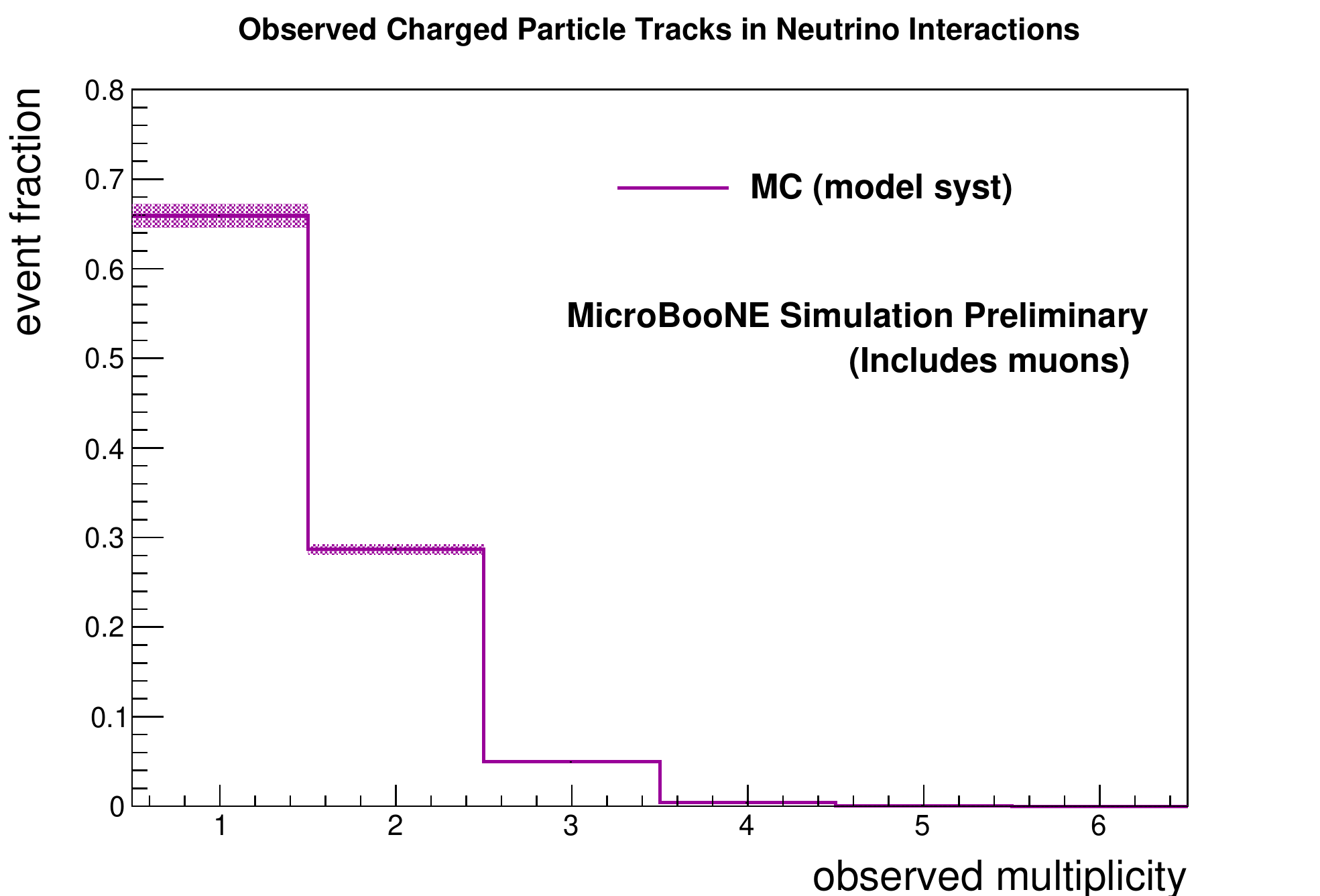}
   \label{img:model_uncer(a)}} \quad
\subfloat[][Background model uncertainty impact; log y scale.]
   {\includegraphics[width=.4\textwidth]{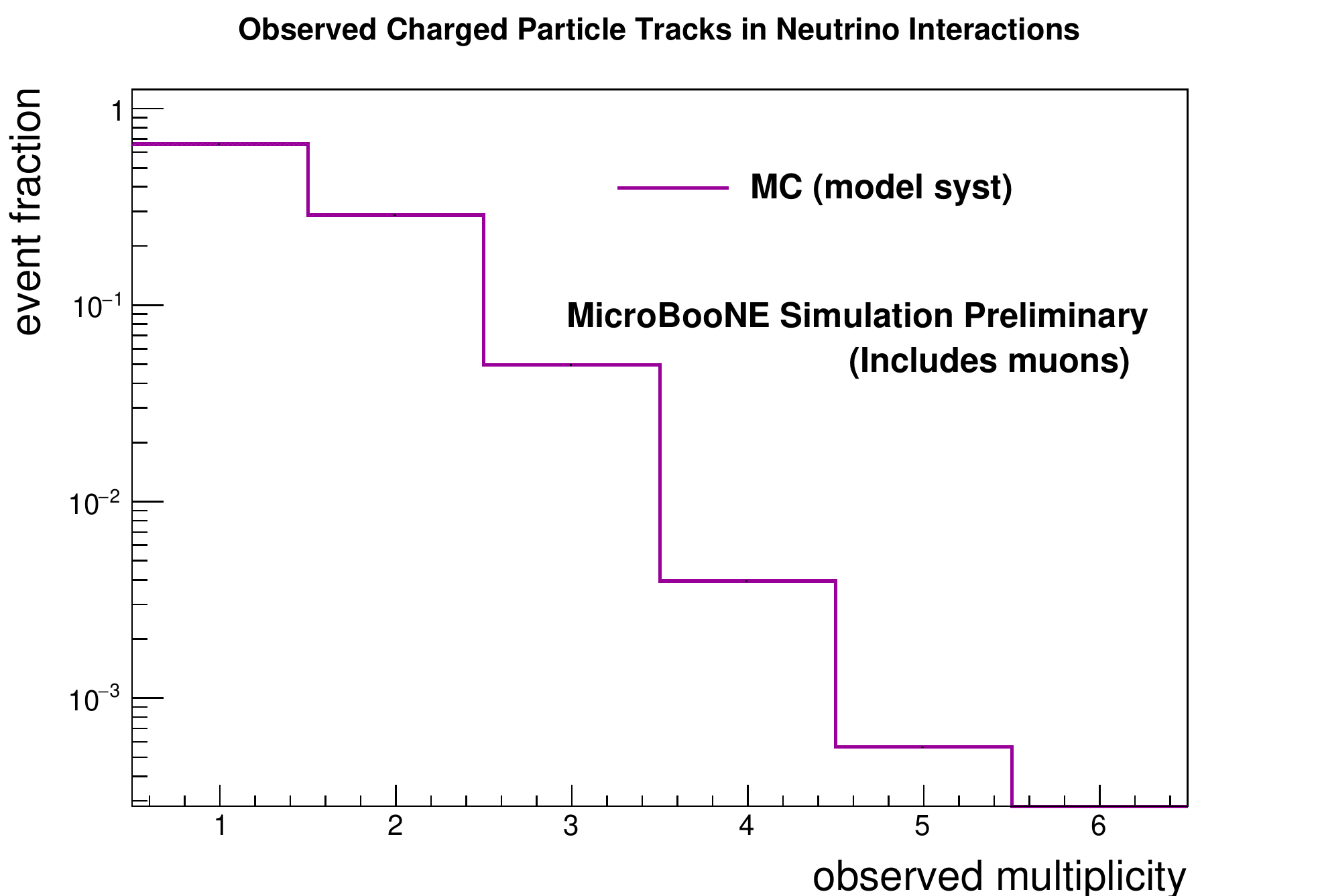}
   \label{img:model_uncer(b)}} \quad

\subfloat[][Flux shape uncertainty impact; linear scale.]
   {\includegraphics[width=.4\textwidth]{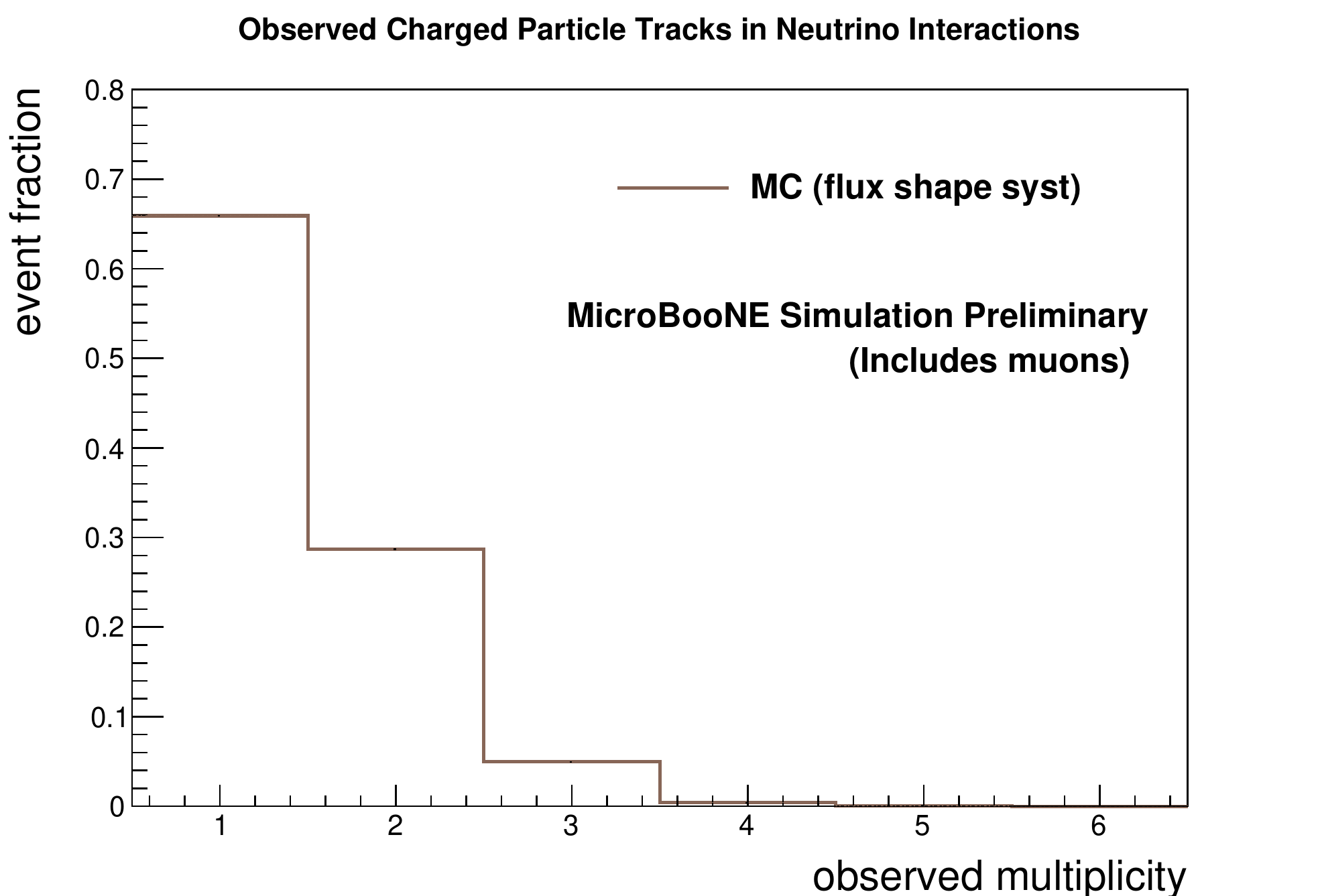}
   \label{img:flux_uncer(a)}} \quad
\subfloat[][Flux shape uncertainty impact; log y scale.]
   {\includegraphics[width=.4\textwidth]{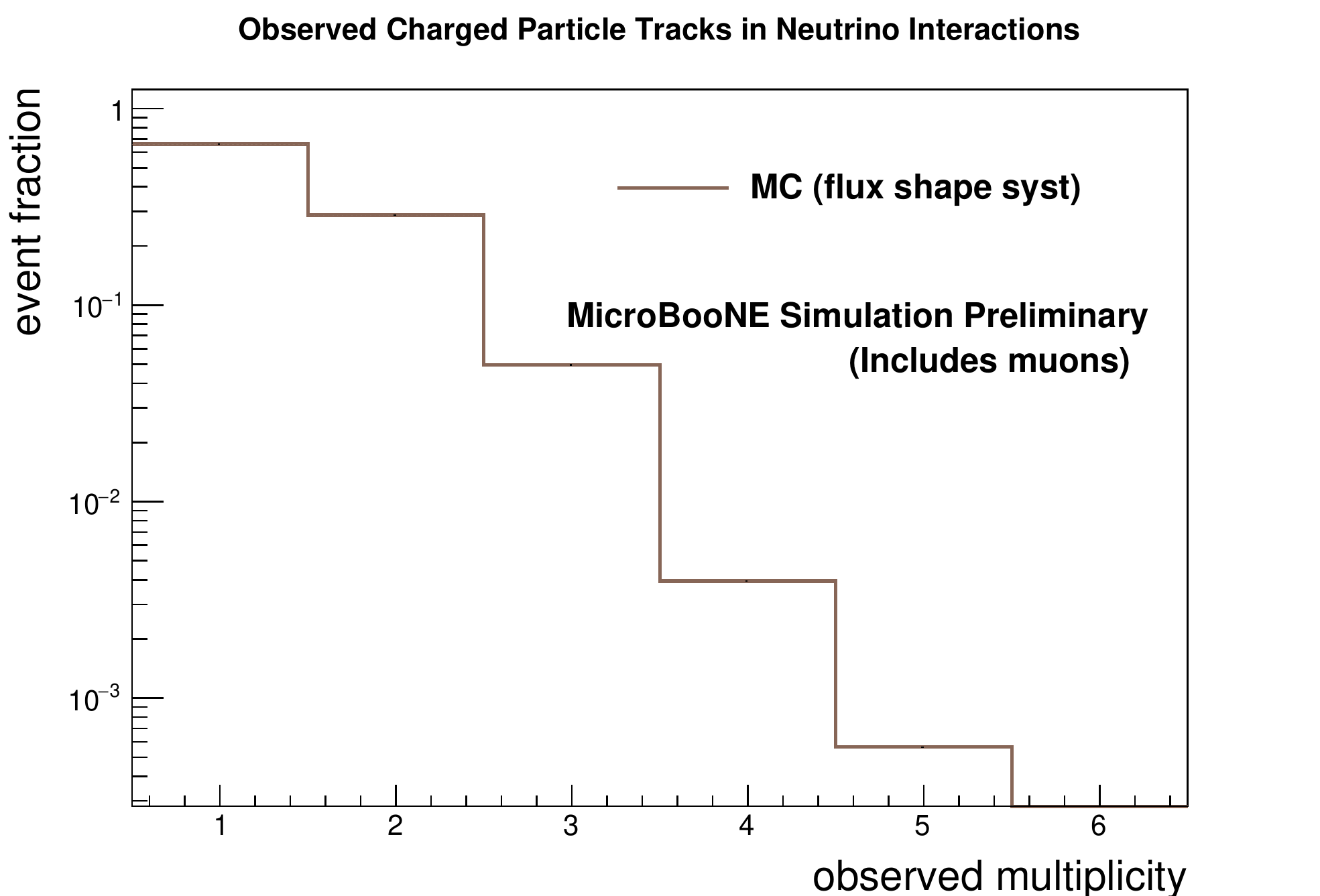}
   \label{img:flux_uncer(b)}} \\
   
\subfloat[][Electron lifetime uncertainty impact; linear scale.]
   {\includegraphics[width=.4\textwidth]{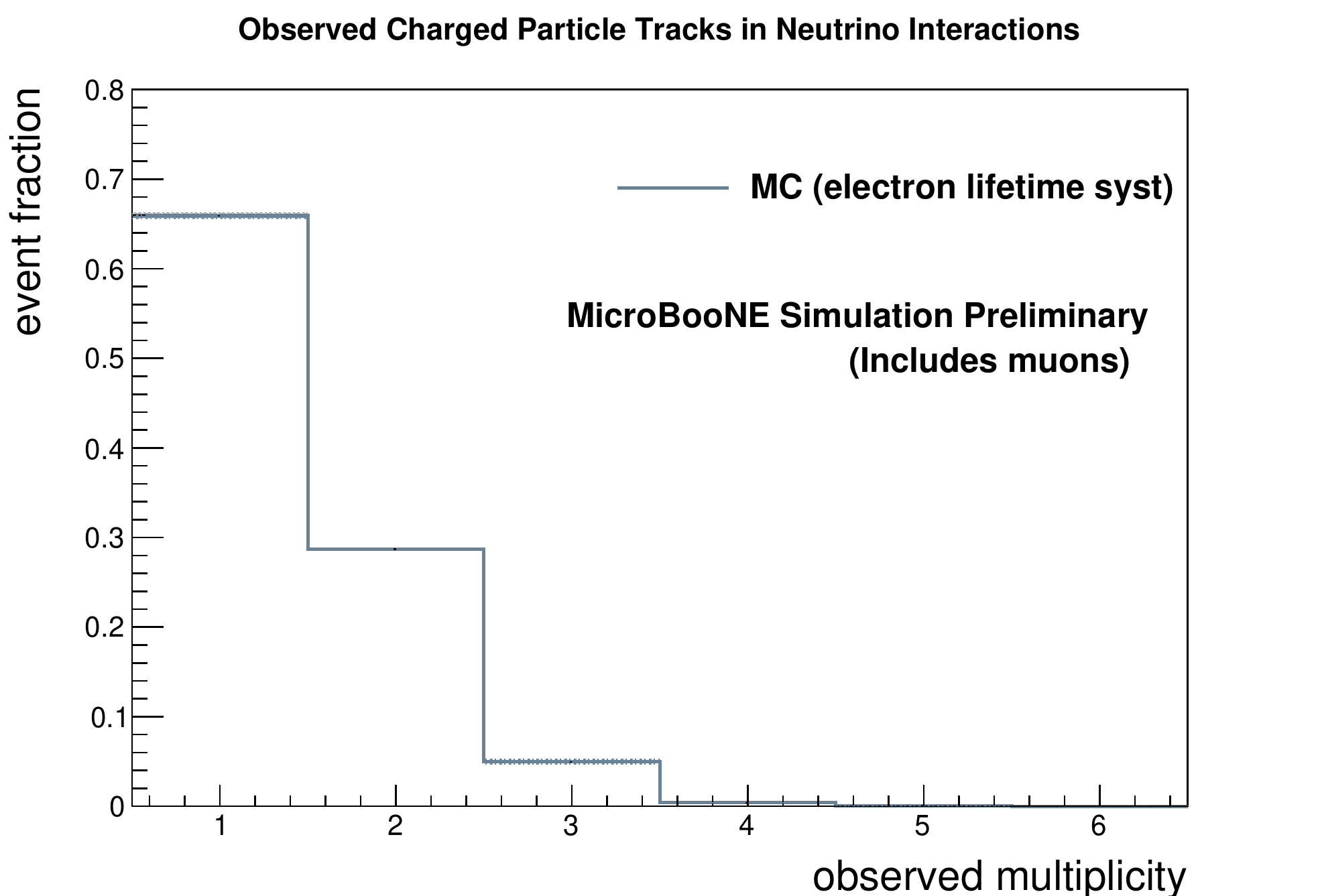}
   \label{img:elec_lifetime(a)}} \quad
\subfloat[][Electron lifetime uncertainty impact; log y scale.]
   {\includegraphics[width=.4\textwidth]{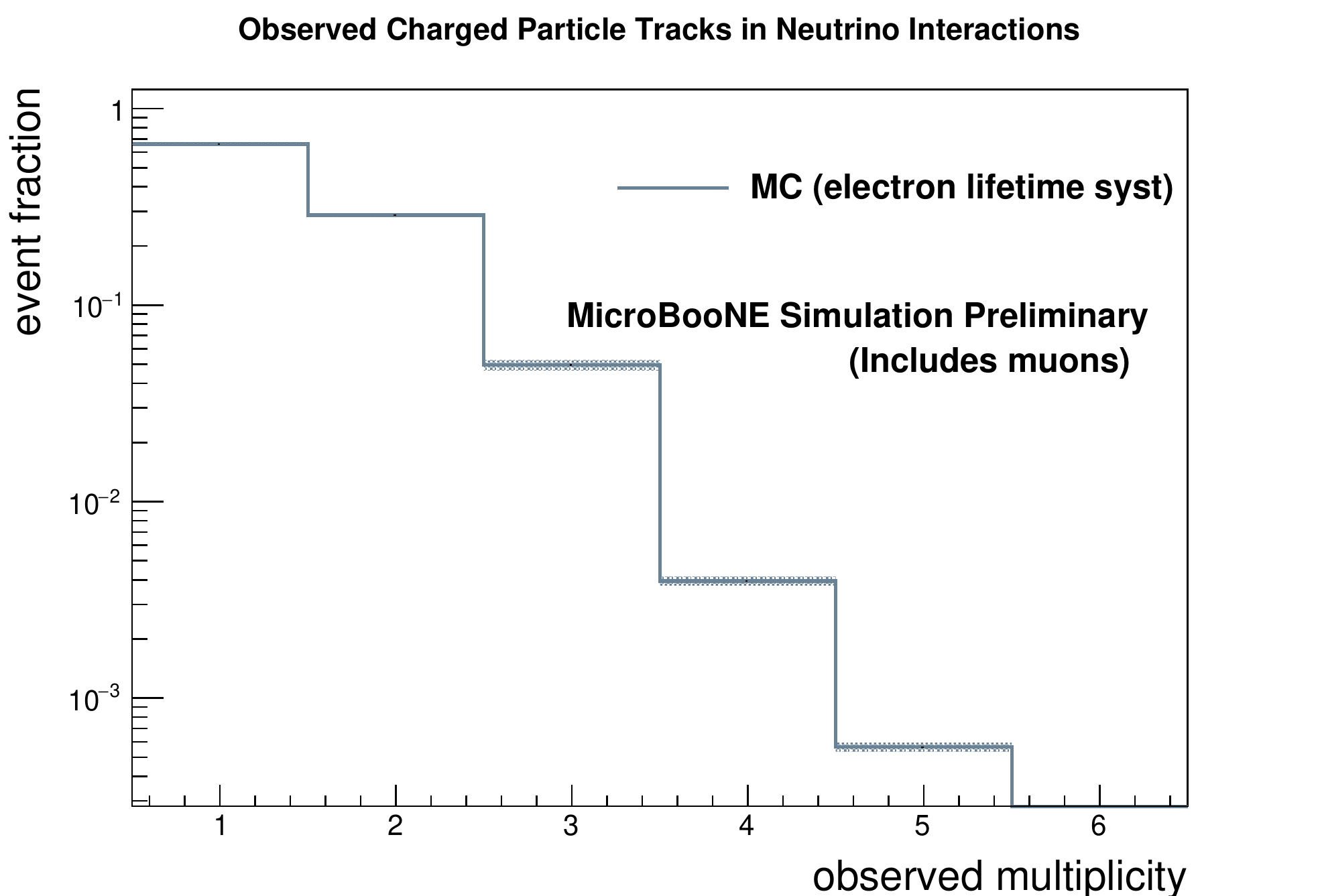}
   \label{img:elec_lifetime(b)}} \\   
   
 \caption{Systematic uncertainty contributions to observed CPMD due to background model uncertainty (top row), flux (middle row), and electron lifetime (bottom row). The width of the line on the histogram indicates the uncertainty band.}
\label{img:sys_effects2}
\end{adjustwidth}
\end{figure}  
   
\subsubsection{Short Track Efficiency Uncertainties}

The dominant systematic uncertainty derives from the analysis of possible differences in the efficiency between data and MC for reconstructing the
shorter length hadron tracks. The overall efficiencies of the
Pandora reconstruction algorithms~\cite{DocDB5987-Pandora-PUB} are a strong
function of the number of hits of the tracks, with plateau not
being reached until of order of several hundred hits. \ The inclusive efficiencies for
reconstructing protons or pions at the 20 CP hit threshold is estimated to
be $\left\langle \epsilon \right\rangle =0.45\pm 0.05$. \ The absolute
efficiency value is not used in this analysis, but we use this estimate to
conservatively assign a mean efficiency uncertainty of $\delta =15\%$.

We then estimate the effect of an efficiency uncertainty on multiplicity by
the following procedure: Consider a track in an event that has made it into a particular multiplicity bin $N$. \ If one lowers the tracking efficiency by the factor $%
1-\delta $, then there is a 1-$\delta$ probability that the track reconstructed and the event stayed in that multiplicity bin, and a probability $\delta $ that the track
would not have been reconstructed and that the event would thus have a \emph{%
lower} multiplicity. \ \ If the overall multiplicity is $N$, with $N-1$ short
tracks and the one long track, and each track's reconstruction probability
is reduced by a factor $1-\delta $, then an overall fraction of events $%
\left( 1-\delta \right) ^{N-1}$ will remain in the bin, and a fraction $%
1-\left( 1-\delta \right) ^{N-1}$ will migrate to lower multiplicity bins.
The fraction of tracks that migrate to multiplicity $N^{\prime }<N$ from bin 
$N$, $f\left( N^{\prime };N,\delta \right) $, is given by binomial
statistics:%
\begin{equation}
f\left( N^{\prime };N,\delta \right) =\frac{\left( N-1\right) !}{\left(
N^{\prime }-1\right) !\left( N-N^{\prime }\right) !}\left( 1-\delta \right)
^{N^{\prime }-1}\delta ^{N-N^{\prime }}.
\end{equation}

We use this result to generate the expected observed CPMD in simulation that
would emerge from lowering the tracking efficiency by the factor $1-\delta $
compared to the default simulated CPMD. \ The difference between the two
distributions is then taken as the systematic uncertainty assigned to short
track efficiency, with the assumption that the effect of increasing the
default efficiency by a factor $1+\delta $ would produce a symmetric change.
Table \ref{Short track efficiency} summarizes this study for the three
GENIE\ models used. \ The observed multiplicity $=1$ bin observed
probability increases because of more \textquotedblleft feed
down\textquotedblright\ of events from higher multiplicity due to the
lowered efficiency, mainly from observed multiplicity $=2$. \ The other
observed multiplicity probabilities accordingly decrease. \ The biggest
effects are in high multiplicity bins because the loss of events from
lowering the efficiency by the factor $\left( 1-\delta \right) $ goes like $%
\left( 1-\delta \right) ^{N-1}$ for multiplicity bin $N$. \ The estimates do
not consider the possibility of \textquotedblleft fake
tracks\textquotedblright\ that could move events to higher multiplicity. Figure \ref{img:eff_short_uncer(a)} and \ref{img:eff_short_uncer(b)} present the short track efficiency error bands on the default simulation.

\begin{table}[tbp]
\caption{
Relative change in observed multiplicity probabilities corresponding to a -15\% uniform reduction in short charged particle tracking efficiencies for three GENIE models:  default, MEC, and TEM.
The missing entry for multiplicity 5 in TEM is due to no event being generated with that multiplicity.}\label{Short track efficiency}%
\centering%
\begin{tabular}{c|c|c|c}

\textbf{Observed multiplicity} & $\frac{\Delta P_{n}}{P_{n}}$\textbf{Default}
& $\frac{\Delta P_{n}}{P_{n}}$\textbf{MEC} & $\frac{\Delta P_{n}}{P_{n}}$%
\textbf{TEM} \\ \hline
$1$ & $+7\%$ & $+7\%$ & $+8\%$ \\ 
$2$ & $-11\%$ & $-12\%$ & $-12\%$ \\ 
$3$ & $-25\%$ & $-25\%$ & $-25\%$ \\ 
$4$ & $-33\%$ & $-36\%$ & $-39\%$ \\ 
$5$ & $-44\%$ & $-48\%$ &  --\\ 
\end{tabular}%

\end{table}%

\subsubsection{Long Track Efficiency Uncertainties}

To first order, the efficiency for reconstructing the $>75$ cm length track
used to define the sample would not be expected to affect the observed
multiplicity distribution, as it is common to all multiplicities and cancels
in the ratio to form observed multiplicity probabilities. \ At second order,
however, a multiplicity dependence could come in that changes the
distribution of observed multiplicity without affecting the overall number
of events. \ A plausible model for this is that higher multiplicity in an
event helps Pandora better define a vertex, and thus helps the event pass
the $\nu _{\mu }$ CC selection filter.

We estimate the size of this effect by comparing the efficiencies obtained
with the Pandora package~\cite{DocDB5987-Pandora-PUB} to simulated
quasi-elastic final states in which both the proton and muon are
reconstructed to charged pion resonance final states in which the proton,
pion, and muon are all reconstructed. \ From this study we conclude that the
efficiency for finding the muon in final states where all charged particles
are reconstructed could be up to $3\%$ higher for charged pion resonance
events (observed multiplicity 3) than quasi-elastic events (observed
multiplicity 2). We then assume, for the purpose of uncertainty estimation,
that this relative enhancement seen for higher observed multiplicity events
in the MC is completely absent in the data.

Table \ref{Long track efficiency} summarizes this study. \ Effects are
generally small compared to those seen in Table \ref{Short track efficiency}%
. \ No dependence on GENIE version is found. Figure \ref{img:eff_long_uncer(a)} and \ref{img:eff_long_uncer(b)} present the long track efficiency error bands on the default simulation.

\begin{table}[tbp]
\caption{
Relative change in observed multiplicity probabilities corresponding to increasing the
 conditional probability for reconstructing the long track by 3\% for each additional track found in the
 event, as suggested by Pandora studies of QE and charged pion resonance production for three GENIE models: default, MEC, and TEM. The missing entry for multiplicity 5 in TEM is due to no event being generated with that multiplicity.}\label{Long track efficiency}
\centering%
\begin{tabular}{c|c|c|c}
\textbf{Observed multiplicity} & $\frac{\Delta P_{n}}{P_{n}}$\textbf{Default}
& $\frac{\Delta P_{n}}{P_{n}}$\textbf{MEC} & $\frac{\Delta P_{n}}{P_{n}}$%
\textbf{TEM} \\ \hline
$1$ & $-1\%$ & $-1\%$ & $-1\%$ \\ 
$2$ & $+2\%$ & $+2\%$ & $+2\%$ \\ 
$3$ & $+4\%$ & $+4\%$ & $+2\%$ \\ 
$4$ & $+7\%$ & $+7\%$ & $+7\%$ \\ 
$5$ & $+9\%$ & $+9\%$ & -- \\ 
\end{tabular}%
\end{table}%

\subsubsection{Background Model Uncertainties}

In the signal extraction fitting procedure, two conditional parameters ($%
\alpha_\nu$ and $\alpha_{CR}$) were extracted from Monte Carlo simulation
and the off-beam data and were kept fixed afterwards. To calculate the
systematic uncertainties on these parameters, their values were varied $\pm
1 \sigma$ of their statistical errors. Those values were propagated in the
observed charged particle multiplicity distribution and the resulting
distributions are shown in Figures \ref{img:model_uncer(a)} and \ref%
{img:model_uncer(b)}. The effect of this systematic was found to be very
small. The systematic errors obtained from different multiplicity bins were
added in quadrature in the final observed charged particle multiplicity
distribution.

\subsubsection{Flux Shape Uncertainties}

\begin{figure}[tbp]
\centering
\includegraphics[width=0.8\textwidth]{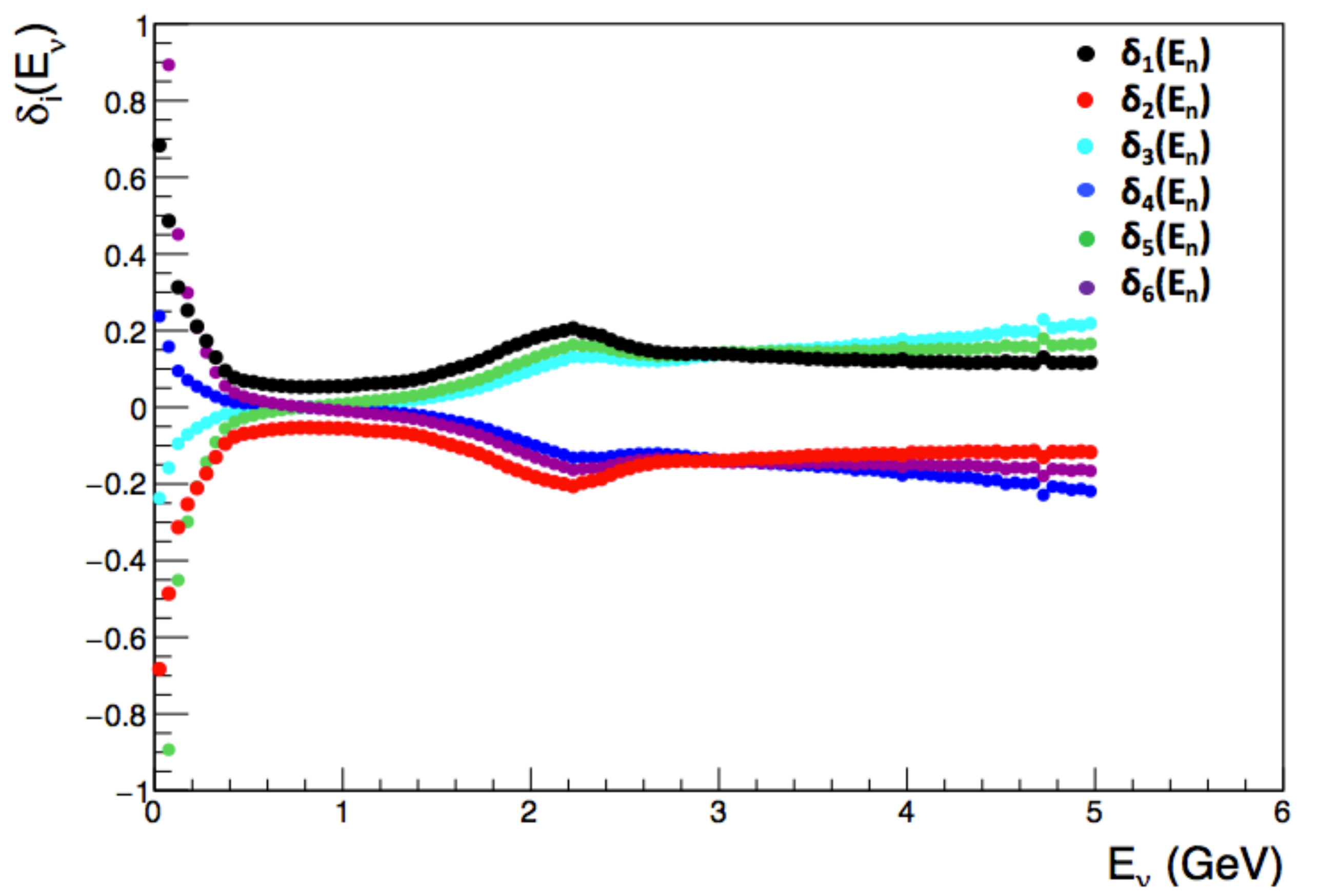} 
\caption{Beam flux shifts for the parameterizations $\protect\delta %
_{i}\left( E_{\nu}\right) $, $i=1-6$. \ The variations $\protect\delta %
_{1}\left( E_{\nu}\right) $ and $\protect\delta _{2}\left( E_{\nu}\right) $
define the envelope of flux uncertainties for the BNB. }
\label{img:flux shifts}
\end{figure}

Variations in flux can be parameterized by%
\begin{equation}
\Phi \left( E_{\nu}\right) \rightarrow \left( 1+\delta \left( E_{\nu}\right)
\right) \Phi \left( E_{\nu}\right) ,
\end{equation}%
where $\Phi \left( E_{\nu}\right) $ is the neutrino flux at neutrino energy $%
E_{\nu}$ and $\delta \left( E_{\nu}\right) $ is the fractional uncertainty in
the flux at that energy. An energy-independent $\delta \left( E_{\nu}\right) $
would have no effect on observed multiplicity distributions as this
measurement is independent of absolute normalization. \ On the other hand,
shifts that, for example, raise the high energy flux relative the low energy
flux could in principle enhance the contributions of higher multiplicity
resonance and DIS\ processes.  We confine ourself to considering highly
correlated energy-dependent shifts, denoted as $\delta _{i}\left(
E_{\nu}\right) $ for $i=1-6$ via an approximate procedure that should be
conservative. \ These shifts, shown in Figure \ref{img:flux shifts} are
allowed to modify the BNB flux within uncertainties determined by the
MiniBooNE\ collaboration~\cite{BNB reference}. \ The first two 
variations simply shift all flux values up ($\delta _{1}\left( E_{\nu}\right) $%
)  or down ($\delta _{2}\left( E_{\nu}\right) $) together according to the
flux uncertainty envelope. \ The next two relatively enhance high energy
flux ($\delta _{3}\left( E_{\nu}\right) $) or  low energy flux ($\delta
_{4}\left( E_{\nu}\right) $) linearly with neutrino energy, with the variation
taken to be zero at the average energy. \ The final two relatively enhance
high energy flux ($\delta _{5}\left( E_{\nu}\right) $) or low energy flux ($%
\delta _{6}\left( E_{\nu}\right) $) logarithmically with neutrino energy,
with the variation taken to be zero at the average energy. As expected,
shifts that are positively correlated across all energies produce negligible
differences, but even shifts that produce sizable distortions between high
and low energies contribute a systematic uncertainty contribution that is
small. Figure \ref{img:flux_uncer(a)} and \ref{img:flux_uncer(b)} present the flux systematic error bands on the default simulation.

\subsubsection{Electron Lifetime Uncertainties}

The measured charge from muon-induced ionization can vary over the detector
volume due to the finite probability for drifting electrons to be captured
by electronegative contaminants in the liquid argon. \ This capture
probability can be parameterized by an electron lifetime $\tau $. We perform
our analysis on simulated data with two lifetimes that safely bound those
found during data taking detector operation conditions, $\tau =$ 6 msec and $\tau
=\infty $ msec . \ Figures \ref{img:elec_lifetime(a)} and \ref%
{img:elec_lifetime(b)} show the result that electron lifetime uncertainties minimally affect this analysis.

\subsubsection{Other Sources of Uncertainty}

A systemic comparison was performed of all kinematic quantities entering
this analysis between off-beam CR data and the CR events simulated with
CORSIKA. No statistically significant discrepancies were observed between
event selection pass rates applied to off-beam data vs. MC simulation.

A check of possible time-dependent detector response systematics was also
performed by dividing the on-beam data into two samples and performing the
analysis separately for each sample. Differences between
the two samples are consistent with statistical fluctuations.

 The data are not corrected for $\nu_\mu$ NC, $\nu_{e}$, $\bar\nu_e$, or $\bar{\nu}_\mu$ backgrounds. An assumption is made that the Monte Carlo simulation adequately describes these non $\nu_\mu$ CC backgrounds.

\subsection{Summary of Uncertainties}

Table \ref{tab:sys_uncer} summarizes the statistical and systematic
uncertainty sources that were studied and our assessment of each of them. \ 

\begin{table}[ptb]
\caption{Statistical and systematic uncertainties estimates. The blank entry
corresponds to having no data statistics in that bin. }
\label{tab:sys_uncer}
\begin{center}
\begin{tabular}{c|c|c|c|c|c}
& \multicolumn{5}{c}{\textbf{Uncertainty Estimates}} \\ 
\textbf{Uncertainty Sources} & \textbf{mult=1} & \textbf{mult=2} & \textbf{mult=3} & \textbf{mult=4}& \textbf{mult=5} \\ \hline
Data statistics & 7\% & 10\% & 38\% & 100\% & -- \\ 
MC statistics & 3\% & 4\% & 7\% & 21\% & 50\% \\ 
Short track efficiency & 7\% & 11\% & 25\% & 33\% & 44\% \\ 
Long track efficiency & 1\% & 2\% & 4\% & 7\% & 9\% \\ 
Fixed model parameter systematics & 2\% & 2\% & 0\% & 0\% & 0\% \\ 
Flux shape systematics & 0\% & 0.4\% & 0.2\% & 0.5\% & 0.8\% \\ 
Electron lifetime systematics & 0.5\% & 0.1\% & 6\% & 5\% & 5\% \\ 

\end{tabular}
\end{center}
\end{table}

\subsection{Experimental Results}

Following the successful implementation and closure test on MC simulation,
the same maximum likelihood fit was performed on MicroBooNE data. Table \ref%
{tab:fit_values_BNB+Cosmic_data} lists the values of the fit parameters
obtained for the data; and Table \ref{tab:mult_fit_data} lists the number of
neutrino events in different multiplicity bins for the MicroBooNE data.
While our method does not require this to be the case, we note that the
fitted PH and MCS test probabilities $P(PH)$, $Q(PH)$, $P(MCS)$, and $Q(MCS)$
agree in data and simulation within statistical uncertainties.

\begin{figure}[tbp]
\centering
\includegraphics[width=0.8\textwidth]{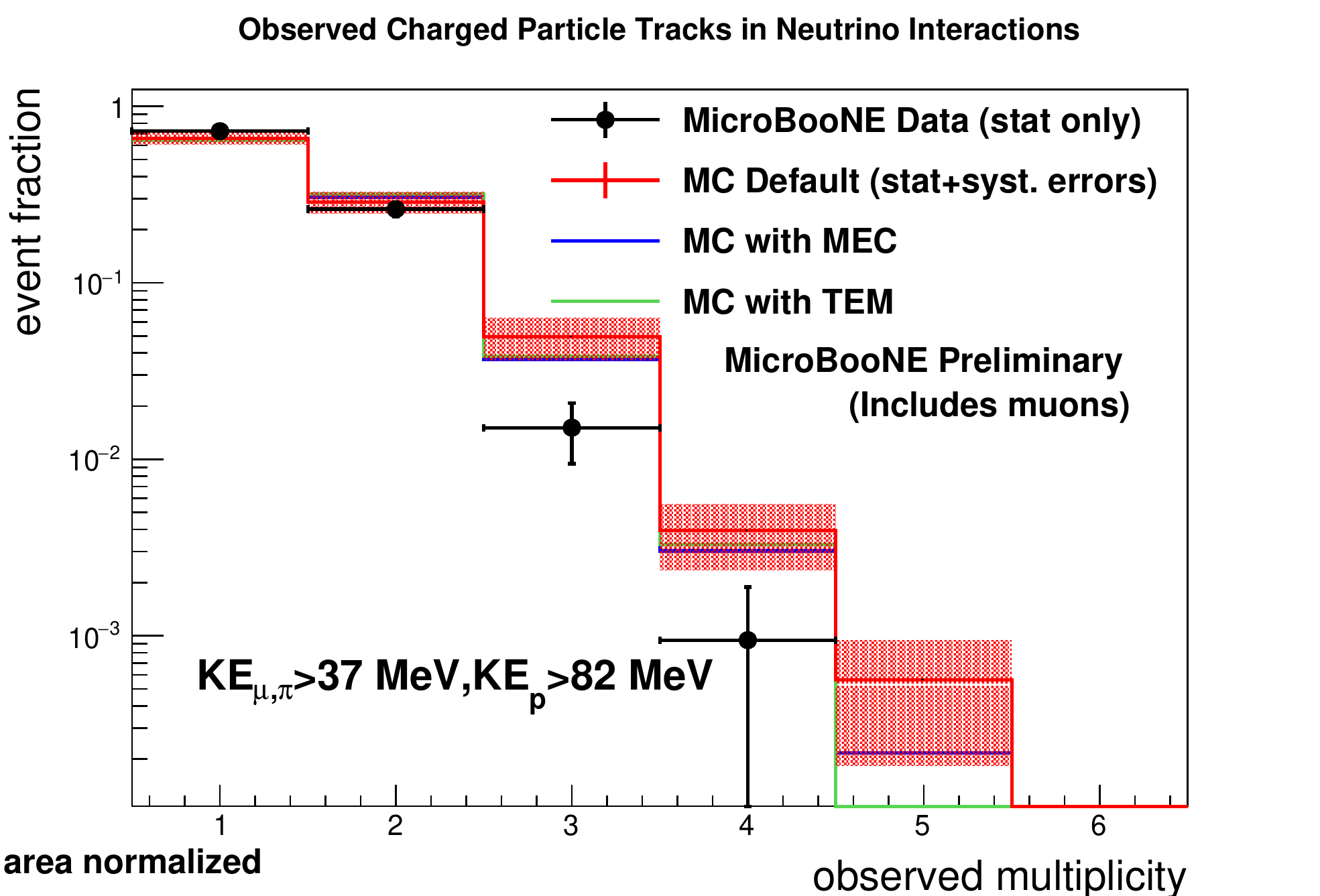} 
\caption{A bin-by-bin fitted, area normalized, CR
background-subtracted, observed neutrino multiplicity distributions for MicroBooNE data overlaid with three GENIE predictions in log y scale. Data error bars include statistical errors
obtained from the fit. Monte Carlo error bands include MC statistical errors from the
fit and systematic error contributions added in quadrature. }
\label{img:final_mult_dist_logy}
\end{figure}

Area normalized, bin-by-bin fitted multiplicity distributions
from three different GENIE predictions overlaid on data are presented in Figure \ref%
{img:final_mult_dist} and \ref{img:final_mult_dist_logy} where data error bars include statistical errors
obtained from the fit and the MC error bands include MC statistical and
systematic errors that are listed in Table \ref{tab:sys_uncer} added in quadrature. Note that the kinetic energy threshold ranges from $37$ MeV for a pion to $82$ MeV for a proton, and this measurement has no acceptance for particles with kinetic energies below these thresholds. Above these minimum acceptance threshold for each particle type, the acceptance curves are rising as a function of particle momentum and angle. However, note that at this stage the measurement has not yet been corrected for non-flat acceptance curves.

In general the three GENIE\ models considered agree within uncertainties
with one another and the data. There are weak indications that the GENIE
models underestimate the number of observed one-track events and
overestimate the number of higher multiplicity events relative to the data.

\begin{table}[tbp]
\caption{Fitted number of neutrino events for the MicroBooNE data sample in
different multiplicity bins. Errors corresponds to the statistical uncertainty estimates obtained from the fit. }
\begin{center}
\begin{tabular}{c|c}
\textbf{Multiplicities} & \textbf{Fit $N_\nu$} \\ \hline
1 & 766$\pm$52 \\ 
2 & 277$\pm$27 \\ 
3 & 16$\pm$6 \\ 
4 & 1$\pm$1 \\ 
5 & 0$\pm$0 \\ 
 
\end{tabular}
\end{center}
\label{tab:mult_fit_data}
\end{table}

\begin{figure}[tbp]
\centering
\includegraphics[width=0.8\textwidth]{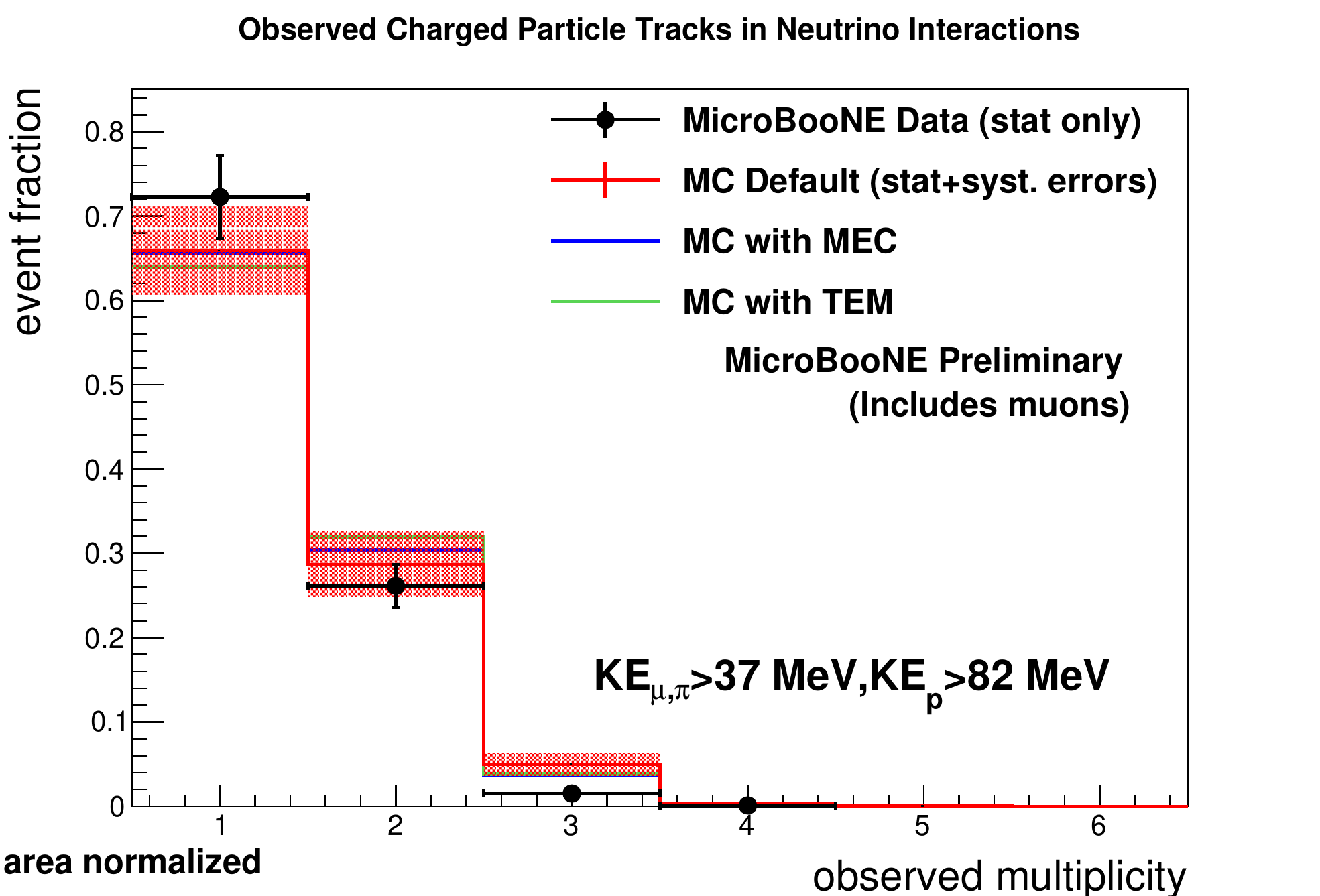} 
\caption{A bin-by-bin fitted, area normalized, CR
background-subtracted, observed neutrino multiplicity distributions for MicroBooNE data overlaid with three GENIE predictions in linear scale. Data error bars include statistical errors
obtained from the fit. Monte Carlo error bands include MC statistical errors from the
fit and systematic error contributions added in quadrature. }
\label{img:final_mult_dist}
\end{figure}

\section{Conclusion and Outlook}

We have completed a preliminary analysis that compares the CR background-subtracted observed charged particle multiplicity distribution in a restricted final state phase space in MicroBooNE data to the
predictions from three GENIE\ tunes processed through the MicroBooNE
simulation and reconstruction. \ Our analysis takes into account statistical
uncertainties in a rigorous manner, and it estimates the impact of the
largest expected systematic uncertainties.

We find all three GENIE tunes to be consistent within uncertainties with the
data, although some weak evidence exists that GENIE under-predicts the
number of one-prong events in the data and over-predicts multiplicities
greater than two. The two alternative GENIE models considered here are not expected to yield significant multiplicity distribution differences; they were chosen to be used in this analysis simply because of their availability, and this analysis serves as proof of principle for future comparisons involving more widely varying model predictions in the future.

As part of this analysis we have developed a data-driven cosmic ray
background estimation method based on the energy loss profile and
multiple Coulomb scattering behavior of muons. \ Within the available Monte
Carlo statistics, we have shown that this method provides an unbiased
estimate of the number of neutrino events in a pre-filtered sample, and
given current statistical precision, it is independent of
signal-to-background level and final state charged particle multiplicity. \
This method could be applied to a broad range of charged current process measurements.

A future publication will use the full data set to reduce statistical
uncertainties, compare to a wider range of neutrino event generator models,
and present the data in a form that corrects for detector acceptance and
efficiency and will be devoted to a measurement of charged particle multiplicity in MicroBooNE over a larger phase space.

\end{document}